\documentclass[usenatbib,useAMS]{mn2e}

\usepackage{natbib}
\usepackage{amsmath}
\usepackage{amssymb}
\usepackage{graphicx}
\usepackage{subfigure}
\usepackage{color}
\usepackage{epsfig}
\bibpunct[,]{(}{)}{;}{a}{}{,}
\allowdisplaybreaks[1]

\newcommand       \be           {\begin{equation}}
\newcommand       \ee           {\end{equation}}
\newcommand       \bea          {\begin{eqnarray}}
\newcommand       \eea          {\end{eqnarray}}

\newcommand{\Ms}{{\rm M}_{\rm shell}}

\def\totd{{\mathrm{d}}}
\def\sun{{\odot}}

\title[Mass Ejection in Failed Supernovae]
{Mass Ejection in  Failed Supernovae: Variation with Stellar Progenitor}
\author[Fern\'andez, Quataert, Kashiyama, \& Coughlin]{Rodrigo Fern\'andez$^{1}$\thanks{E-mail: rafernan@ualberta.ca}, 
Eliot Quataert$^{2}$, Kazumi Kashiyama$^{3}$, and Eric R. Coughlin$^2$\thanks{Einstein Fellow}\\
$^{1}$ Department of Physics, University of Alberta, Edmonton, AB T6G 2E1, Canada\\
$^{2}$ Department of Astronomy \& Theoretical Astrophysics Center, University of California, Berkeley, CA 94720, USA\\
$^{3}$ Department of Physics, University of Tokyo, Bunkyo, Tokyo 113-0033, Japan\\}

\begin{document}

\date{Submitted to MNRAS}

\pagerange{\pageref{firstpage}--\pageref{lastpage}} \pubyear{2017}

\maketitle

\label{firstpage}

\begin{abstract}
We study the ejection of mass during stellar core-collapse when the stalled
shock does not revive and a black hole forms. Neutrino emission
during the protoneutron star phase causes a decrease in the gravitational 
mass of the core, resulting in an outward going sound pulse that steepens 
into a shock as it travels out through the star. 
We explore the properties of this mass ejection mechanism over a range of
stellar progenitors  using spherically-symmetric, time-dependent hydrodynamic
simulations that treat neutrino mass loss parametrically and follow the shock
propagation over the entire star. We find that all types of stellar progenitor
can eject mass through this mechanism.  The ejected mass is a decreasing
function of the surface gravity of the star, ranging from several $M_\odot$ for red
supergiants to $\sim 0.1 M_\odot$ for blue supergiants and $\sim 10^{-3}
M_\odot$ for Wolf-Rayet stars.   
We find that the final shock energy at the surface is a
decreasing function of the core-compactness, and is $\lesssim 10^{47}-10^{48}$~erg 
in all cases. In progenitors with a sufficiently large envelope, high
core-compactness, or a combination of both, the sound pulse fails to unbind
mass.  Successful mass ejection is accompanied by significant fallback
accretion that can last from hours to years.
We predict the
properties of shock breakout and thermal plateau emission produced by the ejection of 
the outer envelope of blue supergiant and Wolf-Rayet progenitors in otherwise
failed supernovae.      
\end{abstract}



\section{Introduction}

Understanding the connection between progenitor stellar properties and remnant
properties after stellar core-collapse has been a longstanding quest in
theoretical astrophysics.  This includes the goal of explaining the observed
neutron star and black hole mass functions (e.g.,
\citealt{ozel_2010,kochanek_2014}).  The origin and properties of stellar-mass
black holes have received renewed attention, and key new empirical constraints,
after the recent detection of gravitational waves from binary black hole
mergers by Advanced LIGO \citep{LIGO16b,LIGO16,LIGO17}. 

Our current understanding of stellar core-collapse indicates that 
black holes can form in a failed explosion or by fallback accretion onto a neutron star
in a successful core-collapse supernova (e.g., \citealt{OConnor_Ott_2011} and
references therein). 
This can occur up to the onset of pair instability in the stellar 
core at initial stellar masses $\gtrsim 150M_\odot$ (e.g., \citealt{kasen_2011} and references therein). 
If the pair instability mechanism succeeds, no remnant is left behind. 
For high enough masses ($\gtrsim 250M_\odot$), however, black holes can 
form promptly since explosive nuclear burning is unable to reverse the 
collapse \citep{fryer_2001}.

Direct observational evidence for black hole formation was reported recently
\citep{adams_2017a} as part of a survey looking for disappearing massive stars
\citep{kochanek_2008,Gerke_et_al_2014}. A weak optical transient reaching $\sim
10^6L_\odot$ and lasting for about a year was observed from a red supergiant
progenitor, with the flux subsequently decaying by a factor of $6$ below the
pre-outburst level. At late-times, the bolometric luminosity decays as
$t^{-4/3}$, which has been attributed to fallback accretion after the collapse
to a black hole.  The detection of such a transient implies that a fraction
$\sim 10\%$ of all core-collapse supernovae result in failures
\citep{adams2017b}.   A significant fraction of failed supernovae can help
explain the absence of massive ($\sim 20-25 M_\odot$) red supergiant
progenitors associated with Type IIp supernovae and the apparent mass gap
between neutron star masses and stellar mass black hole masses
\citep{kochanek_2014}.

The full landscape of observational signatures of black hole formation is not yet well understood, however.
Different scenarios are predicted based on the importance of angular momentum. 
A failed supernova in a star with sufficient rotation
should form a black hole accretion disk, from which transients ranging
from long gamma-ray bursts to fusion-powered explosions can be generated 
(\citealt{bodenheimer_1983,woosley_1993,macfadyen_1999}; see also \citealt{Kashiyama_Quataert_2015}).

If rotation is unimportant,  a transient associated with black hole formation 
in a failed supernova can still be generated
due to the loss of gravitational mass to neutrinos after collapse. Below the onset of pair instability,
black hole formation 
is always preceded by a protoneutron star phase,
in which $\sim 10\%$ of the baryonic mass of the remnant is lost to
neutrino emission (e.g., \citealt{oconnor_2013}). This decrease in
mass generates an outgoing sound pulse that can steepen into a shock and eject the outer 
layers of the star \citep{Nadezhin_1980}.
\citet[ hereafter LW13]{Lovegrove_Woosley_2013} demonstrated this mass ejection effect  
explicitly with time-dependent numerical simulations 
for the case of red supergiant progenitors.  The ejecta gives rise to a transient 
lasting $\sim 1$~yr, with the main
source of power for the light curve being hydrogen recombination.  The predicted properties of these transients are 
broadly similar to the event discovered by \citet{adams_2017a}.   In addition to long time-scale 
recombination-powered emission, \citet{Piro_2013} showed that shock breakout from these events 
should be detectable, and \citet{lovegrove_2017} recently computed the shock breakout signal with 
radiative transfer simulations.

While the neutrino mass loss mechanism can operate in any failed supernova
progenitor, its predictions have only been explored for red supergiants. In particular, 
Wolf-Rayet stars are also strong candidates for black hole formation given the higher compactness of their cores 
(e.g., \citealt{OConnor_Ott_2011,Ugliano_et_al_2012,Sukhbold_Woosley_2014,ertl_2016}).
In addition, our understanding of this mass ejection process is 
based exclusively on the results of numerical simulations.   A better physical description 
of the processes involved would help in understanding all of the routes to black hole formation 
and their observational signatures.

In this paper we explore the physics of the neutrino mass loss mechanism
for a range of stellar progenitors that cover different values of the core compactness
and envelope compactness, including blue supergiants and Wolf-Rayet stars.  
We also provide an analytic derivation of the maximum kinetic energy imparted to 
the outflow from progenitor properties, finding favorable agreement with the results 
of our simulations (a more detailed analytic description of the relevant physics is given 
in \citealt{Coughlin2017}). Finally, we compute the properties of the 
fallback accretion resulting from these models and estimate the observational signatures of the 
weak explosions for different progenitors. 

Our study makes a number of approximations in order to decrease
the computational cost and to efficiently explore parameter space. The main simplification
is the parametric treatment of the loss of gravitational mass by the inner core (following LW13). 
Also, we neglect radiation diffusion, which can alter the dynamics of the ejecta at
late times. Nevertheless, many aspects of the physics, including the propagation of the pressure 
wave inside the star and the transition to an outgoing shock, should be robust and are unlikely to 
change with further improvements in the modeling of the inner core and radiation transport.

The paper is structured as follows. Section~\S\ref{sec:setup} discusses our numerical
methods, including the progenitors chosen, the setup of our hydrodynamic simulations,
the approximations used to model neutrino mass loss, and the parameter range covered
by our models. Section~\ref{sec:energy_scale} contains a derivation of the maximum
kinetic energy imparted to the outgoing shock, and an estimate of the maximum mass
ejected. Section~\ref{sec:results} describes our numerical results, showing first
an overview of fiducial progenitors, analyzing the energetics of mass ejection,
surveying the ejecta properties for various progenitors, discussing failed models,
and analyzing the properties of fallback accretion. Section~\ref{sec:observations}
discusses observational implications, and Section~\ref{sec:sum+dis} summarizes
our results. Appendix~\ref{s:appendix_numerical} shows tests of our numerical
implementation, and Appendix~\ref{s:accretion_appendix} contains a derivation
of our semianalytic approximation to the fallback accretion rate at small radii.

\section{Methods}
\label{sec:setup}

\subsection{Progenitor Models}
\label{sec:progenitor}

We consider non-rotating presupernova stellar models
computed with the MESA stellar evolution code version
6794~\citep{Paxton_et_al_2011,Paxton_et_al_2013,Paxton_et_al_2015,paxton_2017}. 
We generate two sets of models, one with solar metallicity
and another with $Z = 10^{-2}Z_\sun$. Each set has main
sequence masses in the range $12-100$~$M_\sun$. 
Models are generated using the same parameters as in 
\citet{fuller_2015}, particularly their choice of overshoot parameters and 
the use of the `Dutch' wind model \citep{dejager_1988,nugis_2000,vink_2001}. 
In addition, we set  {\tt Zbase} (for opacities) 
equal to the initial stellar metallicity, turn on the velocity for the
entire evolution, and employ the simpler $\alpha$-chain {\tt aprox21} nuclear 
network for advanced burning stages, which significantly shortens execution times.
The contribution of radiation to the photospheric pressure is set to its default value 
for massive stars ({\tt Pextra\_factor = -1}). The inlist files used to generate
all of our progenitors are publicly available.\footnote{\tt https://bitbucket.org/rafernan/bhsn\_mesa\_progenitors}

\begin{table*}
\centering
\begin{minipage}{14cm}
\caption{Properties of the MESA presupernova stellar models used in this study (\S\ref{sec:progenitor}). 
Columns from left to right show model name, ZAMS mass, ZAMS metallicity, mass at core-collapse, 
radius at core-collapse in solar radii and in cm, photospheric luminosity at core-collapse,
effective temperature at core-collapse, type of star at core-collapse (RSG: red supergiant, YSG: yellow
supergiant, BSG: blue supergiant, WR: Wolf-Rayet), core compactness parameter (equation~\ref{eq:compactness}),
and envelope compactness (equation~\ref{eq:global_compactness}). Asterisks indicate the baseline progenitor set.}
\begin{tabular}{lcccccccccc}
\hline
Model & $M_{\rm zams}$ & $Z$      & $M_{\rm cc}$ & \multicolumn{2}{c}{$R_{\rm cc}$} & $L_{\rm cc}$ & $T_{\rm eff}$ & Type & $\xi_{2.5}$ & ${\xi_{\rm env}}$\\
      & $(M_\odot)$    & ($Z_\odot$) & $(M_\odot)$  & ($R_\odot$)  & ($10^{11}$~cm) & ($10^5 L_\odot$) & ($10^3$~K)  &      &    &   \\
\hline
R12z00     & 12  & 1           & 10.0  & 1110    & 770  & 0.9 & 3    & RSG  &  0.15 & 0.009 \\
R15z00$^*$ & 15  &             & 10.8  & 1060    & 740  & 1.3 & 3    & RSG  &  0.24 & 0.010 \\
Y22z00     & 22  &             & 11.1  &  690    & 480  & 2.9 & 5    & YSG  &  0.54 & 0.016 \\
B25z00$^*$ & 25  &             & 11.7  &   96    &  70  & 3.8 & 15   & BSG  &  0.33 & 0.12  \\
W26z00     & 26  &             & 11.9  &   1.1   &  0.8 & 4.0 & 140  & WR   &  0.21 & 10.8  \\
W40z00$^*$ & 40  &             & 10.3  &   0.38  &  0.3 & 5.7 & 260  & WR   &  0.37 & 27.1  \\
W50z00     & 50  &             &  9.2  &   0.42  &  0.3 & 3.4 & 215  & WR   &  0.55 & 21.9  \\
\noalign{\smallskip}                                    
Y25z-2     & 25  & $10^{-2}$   & 23.0  & 940     & 650   & 3.4 & 4.6   & YSG  &  0.25  & 0.024 \\
B30z-2     & 30  &             & 16.0  & 145     & 100   & 5.9 & 13.3  & BSG  &  0.34  & 0.11 \\
B80z-2     & 80  &             & 55.2  &  70     &  50   & 28  & 28.4  & BSG  &  0.97  & 0.79 \\
\hline
\label{t:progenitors}
\end{tabular}
\end{minipage}
\end{table*}

Table~\ref{t:progenitors} shows a sample of models from our complete set, which
covers a representative range in mass, radius, and envelope binding energies. The discussion will
focus on these progenitors, for conciseness. Models are labeled first by the type of star (R: RSG,
Y: YSG, B: BSG, W: WR) and then by their initial Zero-Age 
Main Sequence (ZAMS) mass and metallicity, e.g. R12z00 is the $12M_\sun$ RSG model with solar metallicity.
Models are deemed to have reached the onset of core-collapse when the maximum infall velocity reaches 
$10^8$~cm~s$^{-1}$.

Also shown in Table~\ref{t:models} is the compactness parameter of the stellar core,
which has been shown to correlate well with the onset of BH formation 
\citep{OConnor_Ott_2011,Ugliano_et_al_2012,
Sukhbold_Woosley_2014,Horiuchi_et_al_2014,Pejcha_Thompson_2015}.
We compute this parameter according to
\begin{equation} 
\label{eq:compactness}
\xi_{2.5} = \frac{2.5}{r(M=2.5M_\odot)/1000 \ \rm km}, 
\end{equation}
where $r(M=2.5M_\odot)$ is the radius that encloses a mass $M = 2.5M_\odot$.
Core-collapse simulations that include more physics indicate that the critical value for BH formation 
lies in the range $\xi_{2.5} \simeq 0.2-0.4$, with more compact stars forming
BHs.\footnote{We have also computed the parameters $M_4$ and $\mu_4$ as defined by \citet{ertl_2016},
which can in principle serve as a more stringent predictor of the threshold
for black hole formation. However, using the threshold curves in
\citet{ertl_2016} would predict that all of our models -- including the $12M_\odot$ solar metallicity
model -- should form BHs, even after
computing these two parameters when the central density of stellar models reaches
$5\times 10^{10}$~g~cm$^{-3}$ as in their progenitors.}
The presupernova models in our sample cover this range and higher values
of $\xi_{2.5}$, with the exception of the $12M_\odot$ solar metallicity model,
which is evolved for comparison.

To quantify the degree of binding of the stellar envelope, we also compute a global
compactness at the time of core-collapse, which is simply the mass over the radius
of the star in solar units
\begin{equation}
\label{eq:global_compactness}
\xi_{\rm env} = \frac{(M_{\rm cc}/M_\odot)}{(R_{\rm cc}/R_\odot)},
\end{equation}
where $M_{\rm cc}$ and $R_{\rm cc}$ are the total stellar mass and radius at the onset
of core-collapse (Table~\ref{t:progenitors}). We use $\xi_{\rm env}$ as a proxy for the surface gravity of the star.
Note that the escape speed from the stellar surface is given by $v_{\rm esc} \simeq 600\,\xi_{\rm env}^{1/2}$~km~s$^{-1}$.

\subsection{Numerical Hydrodynamics}
\label{sec:hydro}

We model stellar collapse by solving the time-dependent hydrodynamic equations in 
spherical symmetry using FLASH3 \citep{fryxell00,dubey2009},
\begin{eqnarray}
\label{eq:mass_conservation}
\frac{\partial \rho}{\partial t} + \frac{1}{r^2}\frac{\partial}{\partial r}\left(r^2\rho v_r\right) = 0\\
\label{eq:momentum_conservation}
\frac{D v_r}{D t} + \frac{1}{\rho}\frac{\partial p}{\partial r}
+ \frac{GM(r,t)}{r^2} = 0\\
\label{eq:energy_conservation}
\frac{D e_{\rm int}}{D t} - \frac{p}{\rho^2}\frac{D \rho}{D t} = 0,
\end{eqnarray}
where $\rho$, $v_r$, $e_{\rm int}$, $p$, and $M(r,t)$ are the fluid density,
radial velocity, specific internal energy, pressure, and enclosed mass at
radius $r$, respectively, and $D/Dt \equiv \partial/\partial t +
v_r\partial/\partial r$.  The public version of the code has been modified to
include a non-uniformly spaced radial grid \citep{F12}. The system of equations
(\ref{eq:mass_conservation})-(\ref{eq:energy_conservation}) is closed with the
\emph{Helmholtz} equation of state \citep{timmes2000} and solved with the split
Piecewise Parabolic Method \citep{colella84,fryxell1989}. We do not include
any weak interactions or nuclear burning, since these processes do
not influence the dynamics (other than via neutrino mass loss in the protoneutron
star, which is parameterized; see \S\ref{sec:inner_core}).

The computational domain extends from a minimum radius near the edge of
the iron core, $R_{\rm min} = 2\times 10^8$~cm, to a maximum radius 
$R_{\rm max} = 2\times 10^{16}$~cm, well outside the radius of the largest
red supergiant (RSG) in our sample. Smaller outer boundary radii are taken
for smaller stars.
The grid has logarithmic spacing, with $256$ cells per decade in radius, or equivalently
a fractional cell size $\Delta r / r \simeq 0.9\%$.
Outflow boundary conditions are used at both ends of the computational domain.
All primitive variables are copied from the active cell adjacent to the
boundary into the ghost cells.  The radial velocity has an additional $r^{-2}$
dependence so that the mass flux is constant in the ghost cells. No mass is
allowed to enter the domain.

We implement a remapping procedure to move the inner boundary outward at
specific times in the simulation. The outgoing pressure wave forms within a collapsing medium, 
and most of the material inside this wave falls toward the center at supersonic speeds.
The minimum time step in the simulation is almost always
set by the Courant condition at the inner boundary, where material is
undergoing supersonic infall. In order to speed up calculations, we remove
the innermost decade in radius from the computational domain once this region
has achieved supersonic infall in its entirety and therefore loses causal contact
with the rest of the computational domain (following an approach similar to
\citealt{hammer_2010}). The procedure is first carried out at $t=100$~s and repeated after 
every decade in time, unless conditions make it infeasible (e.g. if a reverse shock modifies 
the velocity field). Since the grid is logarithmic, the minimum time step increases by at 
least a factor ten each time the procedure is repeated. This allows us to follow the evolution of the shock
all the way to the surface of the largest RSGs. More details on this procedure are provided
in Appendix~\ref{s:appendix_numerical}.

\begin{table*}
\centering
\begin{minipage}{16.5cm}
\caption{List of hydrodynamic models evolved and summary of results. 
Columns from left to right show model name, spatial resolution,
type of neutrino mass loss (exp: exponential, full: full loss, max: maximum loss; \S\ref{sec:inner_core}), 
cooling time, time to reach TOV mass, total gravitational mass lost, ejecta mass, total ejecta energy, maximum kinetic energy,
radius at which $t_{\rm ff} = \min(\tau_c,\tau_{\rm tov})$ in the progenitor, analytic energy estimate 
(equation~\ref{eq:dEr}), and analytic ejecta mass estimate (equation~\ref{eq:dM_analytic}). No
ejecta mass or energy is assigned to failed models except for Y25z-2\_m, which ejects some matter but it is clearly bound  (\S\ref{s:failed}).}
\begin{tabular}{lccccccccccc}
\hline
Model    & $\Delta r/ r$  & $\nu$-loss  & $\tau_c$ & $\tau_{\rm tov}$ & $\delta M_{\rm G}$ & $M_{\rm ej}$ & $E_{\rm ej}$ &
           $E_{\rm k,max}$ & $r_c$ & $\Delta E(r_c)$       & $\Delta M$    \\
         &  ($\%$)        &             &  (s)  &    (s)        & ($M_\odot$)     & ($M_\odot$)  & ($10^{47}$~erg) &
           ($10^{47}$~erg) & ($10^9$~cm) & ($10^{47}$~erg) & ($M_\odot$)  \\
\hline
R15z00\_e   &  0.9 & exp  &  3    & 6.1   & 0.30 & 4.2    &  1.5  & 4.7 & 1.5 & 9.9 & 4.8 \\
B25z00\_e   &      &      &       & 3.1   & 0.24 & 4.9E-2 &  1.5  & 4.5 & 1.7 & 12  & 0.18 \\
W40z00\_e   &      &      &       & 2.6   & 0.22 & 5.0E-4 &  0.23 & 3.5 & 1.5 & 11  & 8E-3 \\
\noalign{\smallskip}                                                                     
R15z00\_eHR & 0.45 & exp  &  3    & 6.1   & 0.30 & 4.2    &  1.9  & 4.5 & 1.5 & 9.9 & 4.8 \\
B25z00\_eHR &      &      &       & 3.1   & 0.24 & 4.9E-2 &  1.6  & 4.4 & 1.7 & 12  & 0.18 \\
W40z00\_eHR &      &      &       & 2.6   & 0.22 & 5.0E-4 &  0.25 & 3.4 & 1.5 & 11  & 8E-3\\
\noalign{\smallskip}                                                                 
R12z00\_e  &  0.9 & exp  &  3    & 21    &0.30 & 5.5    & 1.8  & 3.9  & 1.4  & 5.9  & 5.6   \\
Y22z00\_e  &      &      &       & 1.1   &0.12 & ...    & ...  & 0.4  & 0.8  & 4.9  & 1.2   \\
Y25z-2\_e  &      &      &       & 5.3   &0.30 & 2.5    & -1.0 & 8.1  & 1.5  & 19   & 11    \\
B30z-2\_e  &      &      &       &  4    &0.30 & 0.2    & 1.4  & 10   & 1.6  & 19   & 1.1  \\
B80z-2\_e  &      &      &       & 0.2   &0.03 &  ...   & ...  & 0.03 & 0.23 & 1.1  & 0.03  \\
W26z00\_e  &      &      &       & 6.8   &0.30 & 8.1E-3 & 2.6  & 10   & 1.5  & 20   & 0.03\\
W50z00\_e  &      &      &       & 1.2   &0.13 & 5.7E-5 & 0.02 & 0.63 & 0.9  & 5.0  & 5E-3\\
\noalign{\smallskip}                                                                 
R15z00\_f   &  0.9 & full &  3    & 8.0   & 0.47 & 4.6    & 8.8 & 12   & 1.5  & 25   & 4.9   \\
B25z00\_f   &      &      &       & 4.2   & 0.43 & 0.11   & 9.1 & 18   & 1.7  & 40   & 0.30  \\
W40z00\_f   &      &      &       & 3.6   & 0.42 & 4.9E-3 & 3.0 & 17   & 1.7  & 35   & 0.03  \\
B80z-2\_f   &      &      &       & 0.4   & 0.04 & ...    & ... & 0.05 & 0.42 & 1.3  & 0.04  \\
\noalign{\smallskip}                                                                 
R15z00\_m   &  0.9 & max  &  3    & 8.4   & 0.49 & 4.6    & 13  & 17 & 1.5 & 27 & 4.9   \\
B25z00\_m   &      &      &       & 3.7   & 0.37 & 9.5E-2 & 7.0 & 15 & 1.7 & 30 & 0.27  \\
W40z00\_m   &      &      &       & 3.0   & 0.33 & 2.6E-3 & 1.5 & 11 & 1.7 & 22 & 0.02  \\
\hline
\label{t:models}
\end{tabular}
\end{minipage}
\end{table*}

The enclosed mass $M(r,t)$ used to compute the gravitational acceleration in
equation~(\ref{eq:momentum_conservation}) is the sum of the \emph{gravitational} mass
inside the inner boundary $M_G(t)$ and the mass in the computational domain
interior to the radius $r$,
\begin{equation}
M(r,t) = M_G(t) + 4\pi\int_{R_{\rm min}}^{r} x^2\totd x\, \rho(x,t).
\end{equation}
The mass flowing through the inner boundary is added to the \emph{baryonic} mass 
$M_{\rm B}(t)$ inside $R_{\rm min}$,
\begin{equation}
\label{eq:mb_dot}
\dot M_{\rm B} = 4\pi R_{\rm min}^2 \rho(R_{\rm min},t) \max{[-v_r(R_{\rm min},t),0]}.
\end{equation}
Equation~(\ref{eq:mb_dot}) is integrated using the mass flux obtained from the Riemann
solver at the inner boundary, maintaining overall mass conservation close to machine precision
(see Appendix~\ref{s:appendix_numerical}). The gravitational mass $M_G$ is related to the baryonic mass $M_B$ through
neutrino mass loss, which we discuss in the next subsection.

The specific position of inner boundary can have a $\sim 10\%$ effect on the shock energy. 
Appendix~\ref{s:appendix_numerical} presents
test models exploring the impact of this choice, as well as other checks on our numerical
implementation.

\subsection{Evolution of the Inner Core}
\label{sec:inner_core}

Properly modeling the loss of gravitational mass by the protoneutron
star prior to black hole formation would require neutrino radiation hydrodynamic 
simulations in general relativity (as in, e.g., \citealt{OConnor_Ott_2011}). At present there 
is no equation of state that smoothly connects the high-density regime 
required to model the supernova core with the very low density regime
needed to follow the shock beyond the stellar surface, thus an approach that employs
multiple simulation codes would be needed for self-consistent calculations.

Instead, we choose to parameterize the evolution of the inner supernova core by
using  approximations similar to those introduced in LW13.  By default, we use
a simple \emph{exponential} neutrino cooling model, which is a slight variant
of the models considered in LW13.   We assume that 
\begin{equation}
\label{eq:max_loss}
\dot{M}_{\rm G} = \dot{M}_{\rm B} - \frac{\rm BE_{\rm c}(M_{\rm G})}{\tau_{\rm c}}e^{-t/\tau_{\rm c}},
\end{equation}
where $\tau_{\rm c}$ is
a fiducial neutrino cooling timescale, and
\begin{equation}
\label{eq:be_ns}
{\rm BE}_{\rm c} \simeq 0.084 \left(\frac{M_{\rm G}}{M_\odot} \right)^2 M_\odot.
\end{equation}
is the gravitational binding energy of a cold neutron star \citep{lattimer_1989,prakash_1997,lattimer_2001},
obtained as a numerical fit to the relation
\begin{equation}
\label{eq:binding_energy_cold}
{\rm BE}_{\rm c} =  M_{\rm B} - M_{\rm G}
\end{equation}
in cold neutron star models constructed with various equations of state. 

For comparison, we also include the \emph{full loss} model of LW13, which
accounts for the thermal energy stored in the protoneutron star,
\begin{equation}
\label{eq:dotMGh}
\dot M_{\rm G} = \dot M_{\rm B} - (1-\epsilon)\frac{\totd {\rm BE}_{\rm c}}{\totd M_{\rm B}}\dot M_{\rm B} + \dot M_{\rm th}, 
\end{equation}
where the thermal mass-energy equivalent evolves according to
\begin{equation}
\label{eq:dotMth}
\dot M_{\rm th} = - \frac{M_{\rm th}}{\tau_c} + \epsilon \frac{\totd {\rm BE_c}}{\totd M_B} \dot M_B.
\end{equation}
with $\epsilon < 0.5$ a thermalization factor. In equation~(\ref{eq:dotMGh}), a 
fraction $(1-\epsilon)$ of the binding energy of accreted matter is immediately
radiated away in neutrinos, while the remaining fraction is temporarily stored
as thermal energy in the protoneutron star (equation~\ref{eq:dotMth}).
The derivative of ${\rm BE}_{\rm c}$ is obtained from equation~(\ref{eq:binding_energy_cold}):
\begin{equation}
\frac{\totd {\rm BE_c}}{\totd M_B} = \frac{2{\rm BE_c}}{M_{\rm G} + 2{\rm BE_c}}.
\end{equation}

Finally, we also include the \emph{maximum loss} model of LW13
\begin{equation}
\label{eq:maximum_loss_tov}
\dot M_{\rm G} = \dot M_{\rm B} - \frac{\rm BE_{\rm c}(M_{\rm tov})}{\tau_{\rm c}}e^{-t/\tau_{\rm c}},
\end{equation}
which is identical to the default \emph{exponential} model except that the binding
energy in equation~(\ref{eq:binding_energy_cold}) is evaluated at the maximum mass 
of a cold neutron star $M_{\rm tov}$ (corresponding to the Tolman-Oppenheimer-Volkov 
limit, or TOV for short) supported by the equation of state of nuclear matter.

In the \emph{exponential} and \emph{maximum loss} models, neutrino cooling stops when the 
gravitational mass of the star reaches $M_{\rm tov}$. In the \emph{full loss} model, cooling 
stops when $M_{\rm G}-M_{\rm th}$ reaches $M_{\rm tov}$. Beyond this point, we impose 
$\dot M_{\rm G} = \dot M_{\rm B}$.

While more sophisticated models of protoneutron star cooling find a nearly power-law 
evolution of the neutrino luminosity with time instead of an exponential 
(e.g., \citealt{pons_1999,roberts_2012}), we employ the simpler parameterizations of 
LW13 for continuity. In all of our models, we use $\tau_c = 3$\,s, which is
a characteristic timescale over which the neutrino luminosity decreases by
a factor of $\sim 3$ in more detailed calculations. A more thorough
exploration of the effect of this input physics on mass ejection should employ
a neutrino transport code. 

\subsection{Initial Condition and Models Evolved}
\label{s:models_evolved}

For a given presupernova model, the stellar profile data is mapped into FLASH using
interpolation at cell centers. To minimize transients, the mapping uses pressure, density, 
and mass fractions as independent quantities, recovering the remaining thermodynamic variables 
using the Helmholtz EOS. The radial velocity is mapped independently, and the enclosed mass in 
the computational domain is computed from the density field after mapping is complete.

The initial condition for equations (\ref{eq:mb_dot}), (\ref{eq:max_loss}), (\ref{eq:dotMGh}) 
and (\ref{eq:dotMth}) is $M_{\rm G} = M_{\rm B}$ and $M_{\rm th} = {\rm BE}_{\rm c}$, 
with $M_{\rm B}$ equal to the mass enclosed by $r=R_{\rm min}$ in the progenitor star. In all of our models, 
we take $\epsilon = 0.1$, $\tau_c = 3.0 \ \rm s$, and $M_{\rm tov} = 2.5 \ M_\odot$.
The latter is the maximum value allowed by causality constraints (e.g., \citealt{lattimer_2016}).

For numerical reasons, the region outside the star is filled with a constant density ambient medium in hydrostatic
equilibrium. The ambient density is chosen to be 
$\rho_{\rm amb} = \{10^{-18},10^{-16},5\times 10^{-13}\}$~g~cm$^{-3}$ for
RSGs/YSGs, BSGs, and WRs, respectively, with the exception of the $50M_\odot$ WR for which
the ambient density is $10^{-14}$~g~cm$^{-3}$.  These values are such that the
mass in the ambient medium is much smaller than the characteristic ejecta mass over the distances
considered, thus avoiding artificially slowing down the ejecta.   We note, however, that observed WR winds have densities above these values at radii $\lesssim 10 R_\odot$.   If such winds are also present at core collapse, it would modify the shock breakout properties estimated in \S \ref{sec:observations}.

In order to implement these low ambient medium densities, the Helmholtz EOS is extended below its lower density limit of $10^{-10}$~g~cm$^{-3}$ by assuming an ideal gas of electrons instead of its 
standard table for electrons and positrons. The floor of temperature in the simulation is set to 
the lower limit of the Helmholtz table,  $T_{\rm fl} = 10^4$~K, at which hydrogen is still
fully ionized. Simulations are stopped when the temperature inside the shock approaches this floor 
value.

The list of hydrodynamic models evolved is shown in Table~\ref{t:models}. Our default neutrino
mass loss scheme is the \emph{exponential} model (equation~\ref{eq:max_loss}), which we use in
all our progenitors. Model names using this prescription
have `\_e' appended to their names. We adopt three baseline stellar models for more 
detailed study: the $15M_\odot$ solar metallicity RSG (R15z00), the $25M_\odot$ 
solar metallicity BSG (B25z00) and the $40M_\odot$ solar metallicity WR (W40z00). These
three stellar models are evolved at twice our baseline spatial resolution to test
convergence in mass ejection ($\Delta r / r = 0.45\%$), and the corresponding model
names have `HR' appended to them. We also repeat the three baseline progenitors using the
\emph{full loss} model (equation~\ref{eq:dotMGh}) and 
\emph{maximum loss} model (equation~\ref{eq:maximum_loss_tov}). 
These model names have `\_f' and `\_m' appended to them, respectively.

The maximum simulation time depends on the structure of the progenitor. The RSG model is evolved up to
$10^7$~s, shortly after the shock breaks out of the stellar surface. The stopping time is
set by the moment when the fluid behind the shock approaches the floor of temperature in the 
Helmholtz EOS ($10^4$~K), at which point simulations are no longer reliable (in particular,
the internal energy starts to grow and total energy is not conserved). The BSG and WR models
are evolved to $10^6$~s and $10^4$~s, respectively. The criterion for stopping here is 
to achieve nearly constant kinetic and total energies in the ejected shell while at the same time not
sweeping up so much mass in the ambient medium that the shell starts to slow down.

\section{Energy Budget for the Outgoing Sound Pulse and Shock}
\label{sec:energy_scale}

\begin{figure}
\includegraphics*[width=\columnwidth]{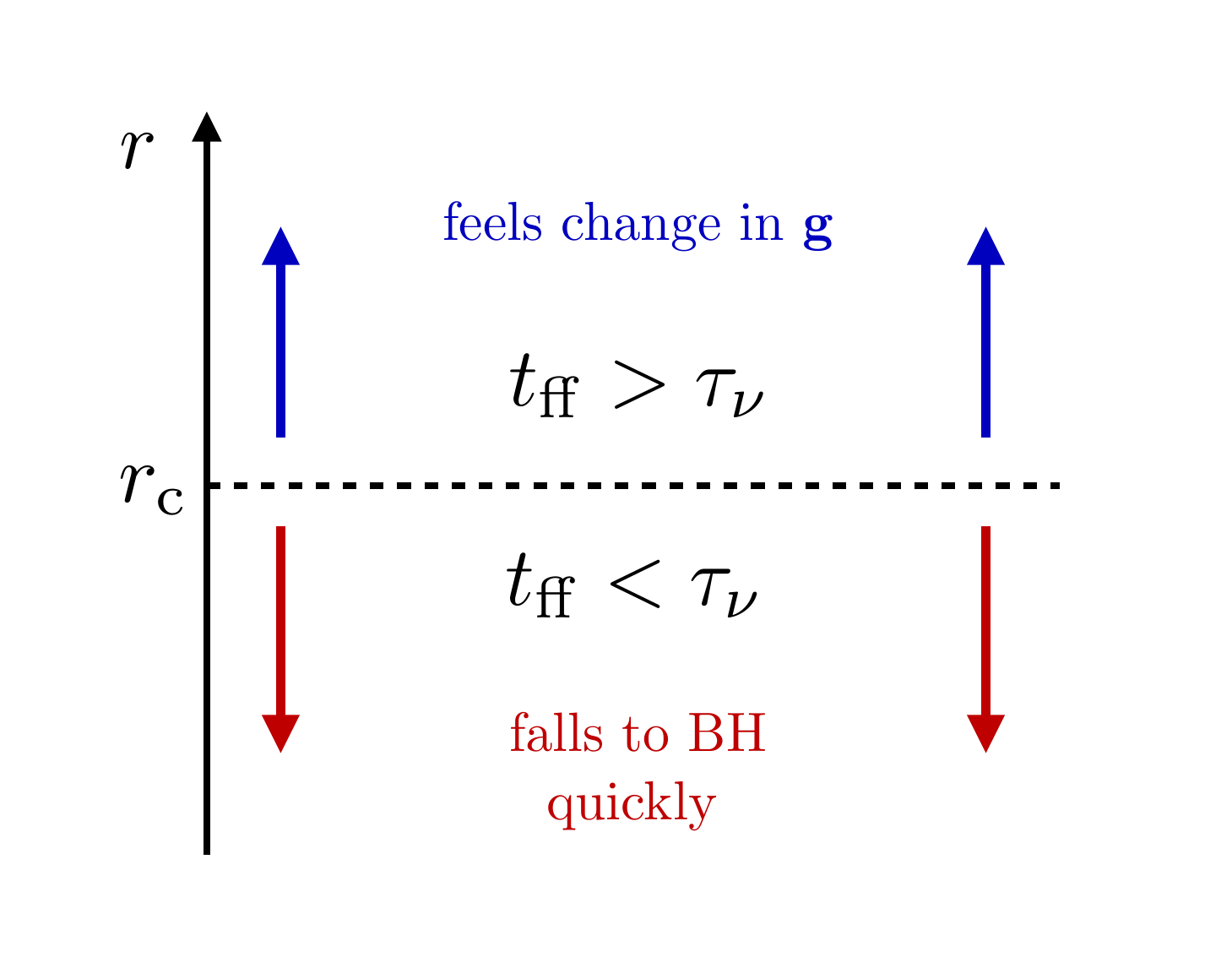}
\caption{Schematic illustration of the location where the outgoing sound
pulse forms in response to the change in gravitational mass, with the vertical 
axis representing the radial coordinate in the 
star. The characteristic radius $r_c$ for sound pulse formation is such that
the free-fall time $t_{\rm ff}$ (equation~\ref{eq:tff}) is equal to the 
time $\tau_\nu$ over which the gravitational mass changes due to neutrino emission.
At small radii, $t_{\rm ff}< \tau_\nu$ and the material falls toward the BH
before experiencing a significant change in the gravitational acceleration. An outgoing 
sound pulse can form in regions that satisfy $r\gtrsim r_c$.  For a wide range of realistic stellar progenitors, $r_c \sim {\rm few} \, 10^9$ cm (Table \ref{t:models}).}  
\label{f:shock_formation_diagram}
\end{figure}

The change in gravitational mass $\delta M_{\rm G}$ due to neutrino emission occurs over 
a finite timescale
\begin{equation}
\tau_\nu = \min\{\tau_c, \tau_{\rm tov}\},
\end{equation}
where $\tau_c$ is the neutrino cooling time in the protoneutron star and
$\tau_{\rm tov}$ is the time to collapse to a BH.
Changes in the gravitational acceleration
act on the different stellar layers on the local free-fall time
\begin{equation}
\label{eq:tff}
t_{\rm ff}(r) = \left[\frac{r^3}{G M(r)}\right]^{1/2},
\end{equation}
which generally is an increasing function of radius. Regions in the
collapsing star for which $t_{\rm ff}\ll \tau_\nu$ fall onto the black hole 
without experiencing any significant change in their gravitational acceleration;
conversely, regions that satisfy $t_{\rm ff}\gg \tau_\nu$ respond instantaneously
to the change in gravity (Figure~\ref{f:shock_formation_diagram}).
The transition between these two regimes lies at a radius $r_{\rm c}$ such that
\begin{equation}
\label{eq:rc_definition}
t_{\rm ff}(r_{\rm c}) =  \tau_\nu.
\end{equation} 
For a wide range of stellar masses, this radius has a characteristic
value $r_{\rm c}\sim 10^9$~cm for a neutrino cooling timescale
of $\tau_\nu \sim 3$~s (c.f. Table~\ref{t:models}).

Once neutrinos change the gravitational mass by $\delta M_{\rm G}$,
the inward gravitational force on each mass shell has decreased and
thus there is a net outward acceleration due to the excess pressure gradient,
with magnitude
\begin{equation}
\label{eq:outward_acceleration}
a = \frac{G\delta M_{\rm G}}{r^2}.
\end{equation}
The outward acceleration produces a net outward velocity over the free-fall time
\begin{equation}
v = a\, t_{\rm ff} = \sqrt{\frac{G M(r)}{r}} \frac{\delta M_{\rm G}}{M(r)}.
\end{equation}
Note that this velocity is everywhere much less than the local escape speed since 
$\delta M_{\rm G} \ll M(r)$. 
\begin{figure}
\includegraphics*[width=\columnwidth]{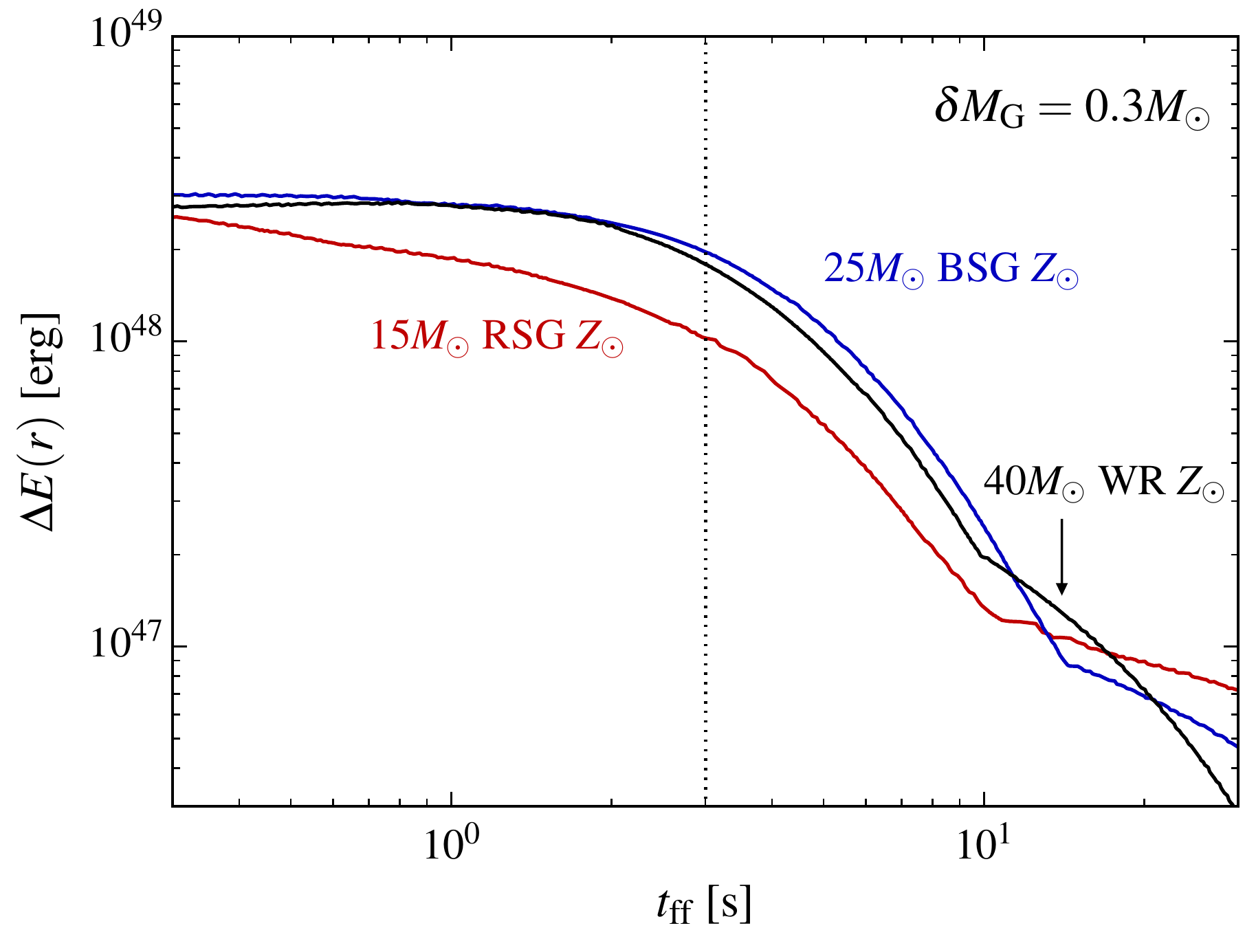}
\caption{Analytic estimate of the kinetic energy in the outgoing sound pulse at radius 
$r$ (equation~\ref{eq:dEr}) as a function of free-fall time (equation~\ref{eq:tff}) 
for the three baseline progenitors, as labeled (c.f. Table~\ref{t:progenitors}). A gravitational
mass loss of $\delta M_{\rm G} = 0.3M_\odot$ has been used. The vertical dotted line marks
a fiducial neutrino cooling time $\tau_\nu = 3$~s, corresponding to radii 
$\{1.5,1.7,1.7\}\times 10^9$~cm in the RSG, BSG, and WR progenitor, respectively.  
The energy input is dominated by radii for which $t_{\rm ff} \gtrsim \tau_\nu$ 
(see Fig. \ref{f:shock_formation_diagram}).}
\label{fig:dE}
\end{figure}

The total energy imparted to a given mass shell by the net 
outward pressure force is given by
\begin{align}
& \Delta E(r)   \simeq  \frac{1}{2} \, \Ms \, v^2 \simeq \alpha \frac{G \delta M_{\rm G}^2}{2 r} \frac{H}{r} \nonumber \\ 
    & \simeq 1\times 10^{48}  \, \left(\frac{\alpha}{0.4}\right) \, \left(\frac{H/r}{0.4}\right) \, 
       \left(\frac{\delta M_{\rm G}}{0.3 \, M_\odot}\right)^2 \, \left(\frac{2\times 10^{9} \, {\rm cm}}{r}\right) \, {\rm erg}
\label{eq:dEr}
\end{align}
where $\Ms =  4 \pi r^2 \rho H = (H/r) dM(r)/d \ln r$ and $\alpha \simeq d \ln
M(r)/d\ln r$, with $H$ the pressure scale height. 
For both super-giant and compact progenitors with $\gtrsim 25
M_\odot$ we have $H/r \sim
0.3-0.5$ and $\alpha \sim 0.4-0.7$ at $r \sim r_c \sim 2 \times 10^9$ cm.

\begin{figure}
\includegraphics*[width=\columnwidth]{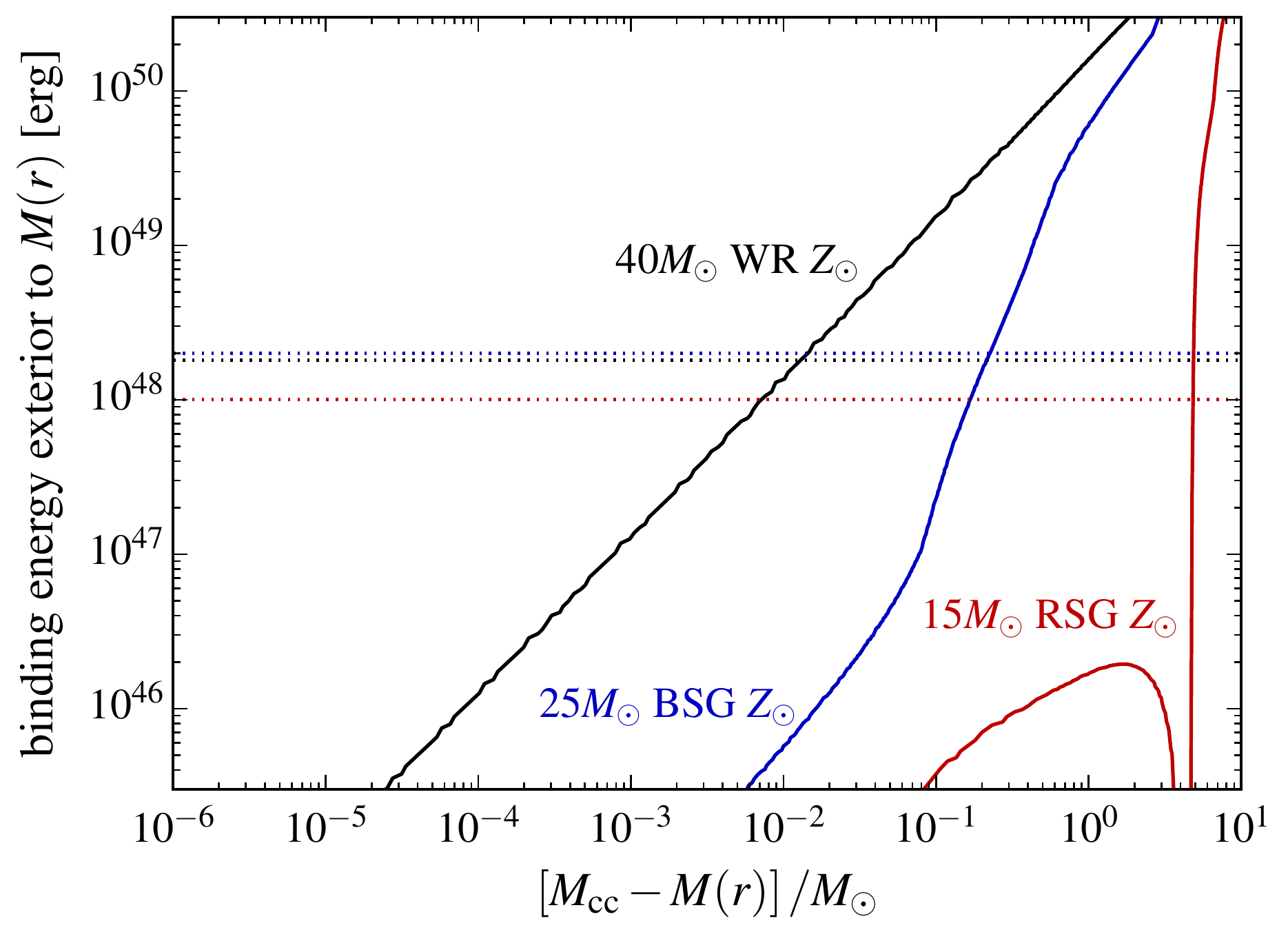}
\caption{Binding energy of the mass exterior to radius $r$ 
for the three baseline progenitors, as labeled (c.f. Table~\ref{t:progenitors}). 
The horizontal dotted lines denote $\Delta E(r_c)$ for each star 
assuming $\delta M_{\rm G} = 0.3M_\odot$ (Figure~\ref{fig:dE}),
and the intersection between curves of the same color yields the upper limit on 
the ejected mass $\Delta M$ for each progenitor (equation~\ref{eq:dM_analytic}):
$\{4.8,0.22,0.013\}M_\odot$ for RSG, BSG, and WR, respectively.
Binding energy changes sign over a narrow radial interval near the base of 
the hydrogen envelope in the RSG progenitor. This is associated with a drop in
the effective adiabatic index $\gamma_{\rm e} = 1 + p/(\rho e_{\rm int})$ below
$4/3$ in the outermost few solar masses of the envelope, overlapping in space
with hydrogen recombination.}
\label{fig:bind}
\end{figure}

Figure \ref{fig:dE} shows $\Delta E(r)$ as a function of the initial free-fall
time at a given radius for our baseline stellar progenitors.
The energy input on the shock is largest at small radii, where the induced outwards 
pressure force is largest. Note, however, that the vertical dotted line in Figure \ref{fig:dE}
is where the free-fall time is equal to $\tau_\nu \simeq 3$~s. Interior to
this radius, the energy injection is strongly suppressed relative to 
equation~(\ref{eq:dEr}) because the mass is incorporated into the neutron star prior to
most of the neutrino binding energy being radiated away. Thus the maximum
energy input into the star by the change in gravitational mass is given by 
equation~(\ref{eq:dEr}) evaluated at radii $\sim r_c$.   \citet{Coughlin2017} present a more detailed derivation of the sound wave excitation, propagation, and energetics induced by neutrino mass loss.  Up to factors of order unity (e.g., $\sim H/r$), their results are consistent with Figure \ref{fig:dE} and our derivation here.

The maximum amount of mass that can be ejected by the shock can be 
estimated by considering the net energy (internal plus gravitational)
of the outermost stellar layers (Figure~\ref{fig:bind}). If the total energy of the shock 
is bounded by $\Delta E(r_c)$, then the maximum amount of mass $\Delta M$ than can be ejected
by the shock is the outermost mass shell with a net energy equal to $\Delta E(r_c)$,
\begin{equation}
\label{eq:dM_analytic}
\Delta E(r_c) = \int_{M_{\rm cc}-\Delta M}^{M_{\rm cc}} (-e_{\rm tot})\,\totd M,
\end{equation}
where $e_{\rm tot}$ is the total specific energy of stellar material (generally negative).  
Figure \ref{fig:bind} shows that $\Delta M$ should be a decreasing function of the
degree of gravitational binding at the surface of the star, which can 
be quantified by the surface gravity or the envelope compactness $\xi_{\rm env}$.

\begin{figure*}
\includegraphics*[width=\textwidth]{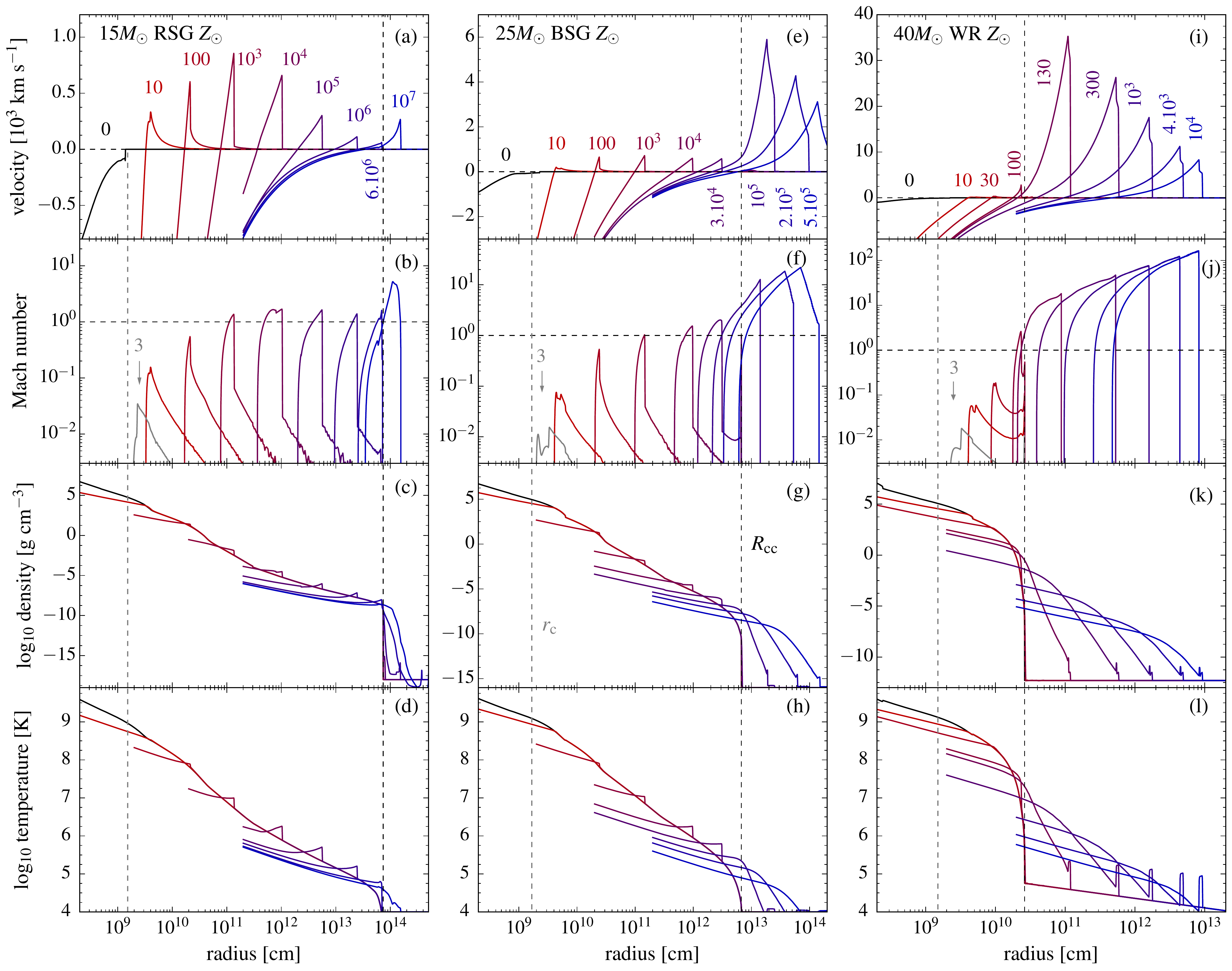}
\caption{Snapshots of radial velocity (top), Mach number of material with positive velocity (second row), 
density (third row) and temperature (bottom) as a function of radius for the three baseline models 
(left column: R15z00\_eHR, middle column: B25z00\_eHR, right column: W40z00\_eHR), with the times in seconds 
as labeled on each curve in the top two rows. 
The black dashed line marks the surface of the star at core-collapse ($r=R_{\rm cc}$), while the gray dashed line denotes the
position of $r_c$ (equation~\ref{eq:rc_definition}). In all models, the inner decade in radius is removed 
from the calculation once infall is supersonic, in order to increase the time step 
(c.f. \S\ref{sec:hydro} and Appendix~\ref{s:appendix_numerical}). The velocity and Mach number of
material initially around and outside the stellar surface (very low density) have been set to zero, 
for clarity.}
\label{f:velx_dens_temp_fiducial}
\end{figure*}


\section{Numerical Results}
\label{sec:results}

\subsection{Overview of Sound Pulse \& Shock Evolution}
\label{s:overview}

In the absence of a change in the gravitational mass due
to neutrino emission, the star simply collapses from the
inside out and the velocity is everywhere negative, with 
the possible exception of the regions outside the stellar surface 
(see Appendix~\ref{s:appendix_numerical} for numerical tests involving pure collapse).

When neutrino mass loss operates, a pressure wave is driven due
to the change in gravitational acceleration. 
Figure~\ref{f:velx_dens_temp_fiducial} shows snapshots of the
radial velocity, Mach number, density, and temperature in the three 
fiducial progenitors. 

The outgoing wave forms as a sound pulse in the vicinity of $r=r_c$ over
a timescale $\tau_\nu = \min (\tau_c, \tau_{\rm tov})$ (c.f. \S\ref{sec:energy_scale}).
Initially, this pulse is subsonic, but as it propagates out, its
leading edge gradually steepens into a shock. 
Three features are noticeable from the velocity and Mach number snapshots in 
Figure~\ref{f:velx_dens_temp_fiducial}. First, once the Mach number exceeds
a value $\sim 0.1$, the outgoing pressure wave is
bound by leading and trailing edges that can be clearly defined. Second, while the 
leading edge eventually becomes supersonic, its Mach number is only slightly
larger than unity while inside the star.
Finally, note also that the entire portion of the star outside the wave
acquires positive velocity, as implied by equation~(\ref{eq:outward_acceleration}).

Upon reaching the stellar surface, the Mach number at the leading edge of
the shock increases, as does the distance between leading and trailing edges.
This behavior is expected from the fact that the density and sound speed in the star decrease
very steeply with distance from the stellar surface, leading to
acceleration of the shock (e.g., \citealt{matzner_1999}).

The formation of the pressure wave is similar in all
stellar progenitors, hence the evolution at times $t \lesssim 10$~s
is qualitatively and even quantitatively similar in all cases. This is expected given 
that the interior structure near $r_c$ is similar in different
models at the onset of core-collapse.
Noticeable differences appear once the
wave propagates into the stellar envelope. In the RSG case, the
shock velocity reaches several $100$~km~s$^{-1}$ throughout the stellar interior,
with no significant increase upon breakout from the
stellar surface. In the BSG model, upon shock breakout the velocity of most of the 
mass jumps to a few $1000$~km~s$^{-1}$, while in the WR case the velocity can reach a few $10^4$~km~s$^{-1}$. The
general trend is therefore larger shock breakout velocities
for increasing envelope compactness $\xi_{\rm env}$.

Note, however, that despite having converged in mass ejection with resolution, 
our fiducial models are not fully resolving the outermost layers of the star and therefore do not fully capture shock acceleration during breakout. The gas pressure scale 
heights at the photosphere in the fiducial RSG, BSG, and WR progenitors are
$H_{\rm phot}/R_{\rm cc} = 0.01$, $0.02$, and $0.002$, respectively, while our highest resolution
models have $\Delta r/r \simeq 0.005$, thus barely resolving RSG and BSG photospheres 
and under-resolving WR surfaces. This likely accounts for the behavior of the final ejecta energies 
in Table~\ref{t:models}, which increase by $\sim 10\%-30\%$ when doubling the resolution
in the fiducial models.

\begin{figure*}
\includegraphics*[width=\textwidth]{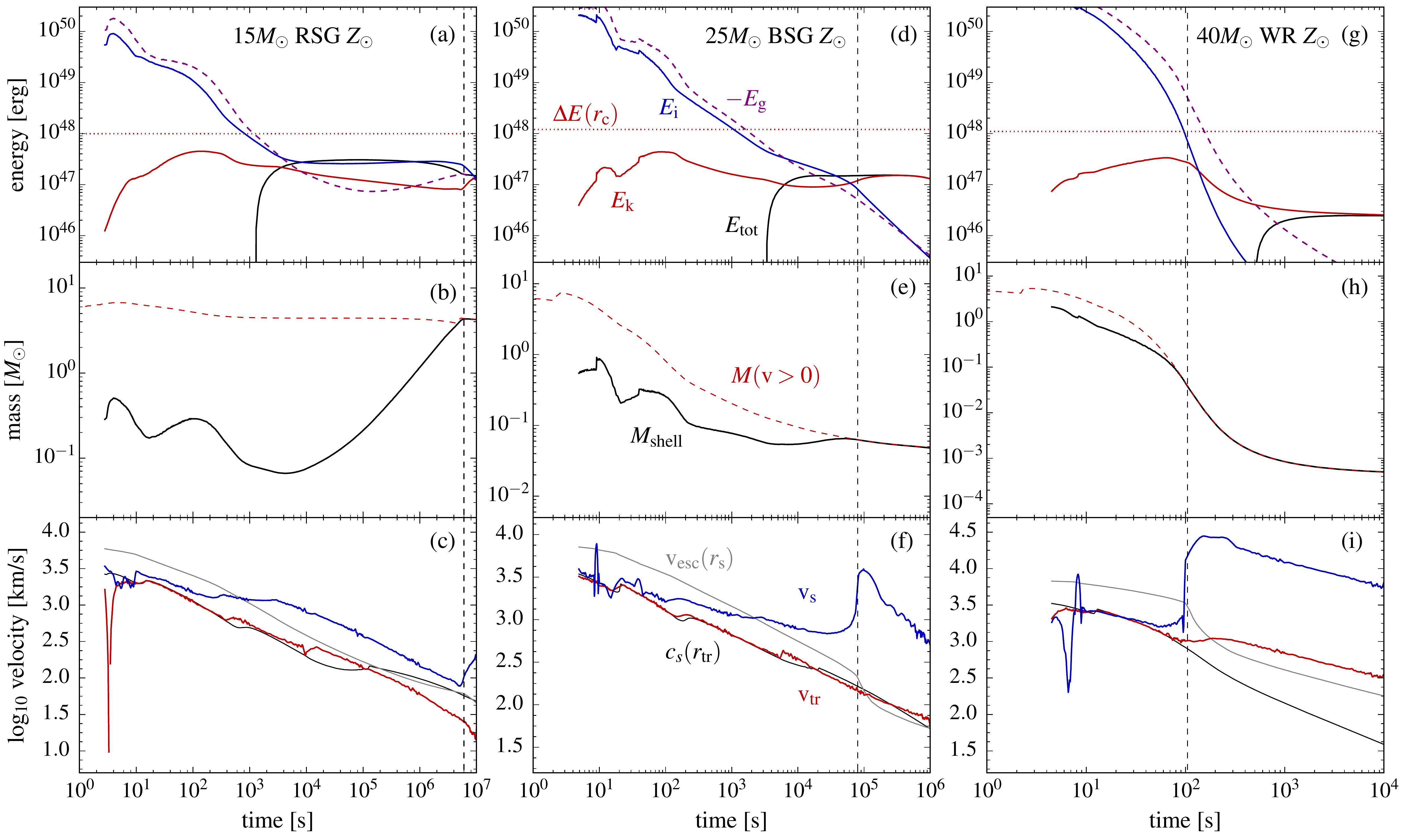}
\caption{Evolution of the energy, mass, and velocity in the outgoing 
pressure wave for models R15z00\_eHR (left column), B25z00\_eHR (middle column), 
and W40z00\_eHR (right column). Tracking of the leading and trailing edge of the wave, 
$r_{\rm sh}$ and $r_{\rm tr}$ respectively (\S\ref{s:overview}), begins once the maximum Mach number exceeds 
$0.03$, for clarity. The vertical dashed line denotes the time at which the leading edge of the
pressure wave reaches the stellar radius $R_{\rm cc}$ (i.e., shock breakout).
\emph{Top row:} Kinetic energy $E_{\rm k}$, internal energy
$E_{\rm i}$, gravitational energy $E_{\rm g}$, and total energy $E_{\rm tot}$
in the wave. The horizontal dotted line shows the analytic estimate of the maximum 
kinetic energy of the shock (equation~\ref{eq:dEr}) evaluated at a radius $r_c$
where $t_{\rm ff}(r_c) = \min(\tau_c,\tau_{\rm tov})$. \emph{Middle row:} Mass contained
in the outgoing pressure wave (solid black). The red dashed curve shows the total mass with
positive velocity, excluding ambient medium. \emph{Bottom row:} radial velocity
of the leading and trailing edges of the wave, $v_{\rm s}$ and $v_{\rm tr}$ respectively. 
Also shown are the local sound speed at the trailing edge $c_s(r_{\rm tr})$, and the local escape
speed at the leading edge $v_{\rm esc}(r_{\rm s})$. The edge velocities are smoothed 
with a Savitsky-Golay filter to suppress noise from numerical differencing.}
\label{f:ener_mass_vshock_fiducial}
\end{figure*}

The radial velocities of the leading and trailing edges of the
wave are shown in the lower row of Figure~\ref{f:ener_mass_vshock_fiducial}.
The trailing edge $r_{\rm tr}$, defined as the point at which the velocity
changes sign, propagates at a speed very close to the local 
sound speed. 
The leading edge $r_{\rm sh}$ moves at a speed slightly faster than 
the trailing edge, with the speed difference increasing 
in magnitude as the leading edge travels into the low density stellar envelope.
The speed of the leading edge eventually exceeds the local escape speed,
either deep inside the star (RSG and BSG) or very near the surface as
it accelerates (WR).

Given the relative weakness of the outgoing shock, the jump in density 
and temperature is small while the outgoing wave is inside the star. 
In particular, a shock develops only in regions
where the temperature is lower than $10^9$~K, hence explosive nuclear burning is not expected. 

For the two RSG cases studied, we obtain shock velocities
of the order of a few $100$~km~s$^{-1}$ and ejecta masses of the order
of a few solar masses, in reasonable agreement with the results of 
LW13.

\subsection{Energetics}

The evolution of the energies contained within the pressure wave are shown in 
Figure~\ref{f:ener_mass_vshock_fiducial} for the three fiducial progenitors.   
These are calculated by integrating over radii between the leading and trailing edges.
In all cases, the kinetic energy $E_{\rm k}$ initially increases, reaching a maximum 
value that is close to (but smaller than) the analytical estimate $\Delta E(r)$ evaluated
at the radius $r_c$ where $t_{\rm ff} \simeq \tau_\nu$. 

The weakness of the shock while deep inside the star is expected given that its energy
is much smaller than the local thermal energy \citep{Tan+01}. In hydrostatic 
equilibrium, the latter is close to the gravitational binding energy. For BSG and WR progenitors,
the characteristic binding energy is $\sim 10^{51}$~erg for most of the stellar interior,
whereas in RSGs the more weakly bound H envelope (Figure~\ref{fig:bind}) provides conditions 
for a shock to develop deeper in the star. The existing theory of shock propagation and 
breakout in stellar interiors \citep{sedov_1959,sakurai_1960,chevalier_1976,chevalier_1982,nadyozhin_1985,
chevalier_1989,kazhdan_1992,matzner_1999} 
assumes a strong shock in which gravity is negligible, which is not applicable in this problem
except when the shock is very close to the stellar surface.

The internal and gravitational energies of the pressure wave, $E_{\rm i}$ and $E_{\rm g}$ respectively,
are initially much higher than the kinetic energy. As the wave propagates out, 
$E_{\rm i}$ and $E_{\rm g}$ decrease
in magnitude. The detailed interplay between internal, gravitational, and kinetic
energy after the maximum in $E_{\rm k}$ depends on the structure of the progenitor.

For the RSG model (R15z00\_eHR), the wave acquires positive total energy 
(black line in Fig.~\ref{f:ener_mass_vshock_fiducial}a) upon 
reaching the base of the hydrogen envelope at time $t\sim 10^3$~s. This position
coincides with a steep radial drop in the binding energy of the outermost layers
of the star (Figure~\ref{fig:bind}).
By this time the leading edge has steepened into a shock with Mach number $\sim 1.5$.
The internal energy stabilizes thereafter, and the gravitational energy reaches a minimum,
increasing afterward as the shock sweeps up mass. Before reaching the stellar surface, 
the internal and gravitational energies are higher than the kinetic energy.
After shock breakout, the kinetic energy quickly increases to the point
at which it matches $E_{\rm i}$ and $E_{\rm k}$. Since the fluid behind
the shock reaches the floor of temperature in the EOS at this point ($10^4$~K),
the subsequent numerical evolution is unreliable and we do not show it. But
assuming that a fraction of the remaining internal energy is used up
in adiabatically expanding the shell, we infer that the asymptotic
energy of the shell should remain within a factor of a few of the value
quoted in Table~\ref{t:models} ($\sim 2\times 10^{47}$~erg, measured
at $t=10^7$~s).

For both the BSG and WR models (B25z00\_eHR and W40z0\_eHR, respectively), 
the internal and gravitational energies decrease almost continuously from the time the
wave forms. The kinetic and total energies become roughly constant once 
$|E_{\rm g}|\lesssim E_{\rm k}$. The key difference between the BSG and WR models
is that in the former the shock acquires positive total energy inside the star,
while in the latter it does so only at several stellar radii from the surface.

The position where the entire outgoing wave becomes unbound 
correlates with the amount of mass in the final unbound shell 
(shown in the middle row of Fig.~\ref{f:ener_mass_vshock_fiducial}).
For the RSG model, the shock becomes unbound deep inside the star, and sweeps
up significant mass in the H envelope. For the BSG and WR models, 
the mass in the shell deceases continuously until positive energy
is achieved, at which point the shell mass stabilizes. The decrease in
the mass in the shock is caused by fallback of the innermost layers of 
the shell that do not become unbound from the star. 

Note also that the entire portion of the star outside $r=r_c$
initially acquires positive velocity (Figure~\ref{f:velx_dens_temp_fiducial}). 
This occurs because the material in this region feels an instantaneous
decrease in the acceleration of gravity (equation~\ref{eq:outward_acceleration}), but it is slower to respond
because the free-fall time is long compared to the formation and
propagation of the shock through the inner layers. While the amount
of mass involved is substantial (Figure~\ref{f:ener_mass_vshock_fiducial}), most of 
this material ends up falling back toward the center since it remains 
gravitationally bound. \citet{Coughlin2017} predict that a second shock can in
some cases emerge at the stellar surface due to this outward motion of the
stellar envelope, particularly in more compact progenitors.  Our simulations do
not fully resolve the layers near the stellar surface to capture this effect
(\S\ref{s:overview}).

In the case of a strong shock propagating through a power-law density medium, two effects compete: the sweeping
up of mass, which slows the shock down, and the pressure gradient
behind the shock, which speeds it up \citep{sedov_1959,herant_1994}. 
Which of these dominates depends on the radial steepness of the density profile. 
For the problem at hand, the shock is not strong enough for gravity to be unimportant, 
hence we may also need to consider the conversion of internal and kinetic energy into 
gravitational potential energy as another source of deceleration.   Except near the 
stellar surface, the leading edge of the shock is
constantly decelerating as it propagates out in all three fiducial 
progenitors (Figure~\ref{f:ener_mass_vshock_fiducial}). This occurs even as the shock sweeps up mass (RSG) or
loses mass to fallback (WR). We thus conclude that the low energy
of the shock makes the effect of gravity the dominant factor
determining the speed of propagation of the shock inside the star.
It would be valuable to understand this regime of shock propagation in 
more detail analytically.

\subsection{Dependence on Neutrino Radiation Model}

Table~\ref{t:models} also shows models obtained by evolving
the fiducial progenitors with the \emph{full loss} and \emph{maximum loss}
prescription for the inner core evolution (\S\ref{sec:inner_core}). 
These models serve to quantify the overall uncertainties in our
results due to the approximate treatment of the protoneutron
star evolution.

In all cases, models lose more gravitational mass than their
counterparts that employ the \emph{exponential} prescription.
While the \emph{maximum loss} prescription results in a larger
decrease in the gravitational mass than the \emph{full loss}
model for the RSG progenitor, the opposite result is found
for the BSG and WR stars. Differences arise in part from the
times needed to reach the TOV limit
and in part by the rate at which mass is radiated away.

Table~\ref{t:models} shows that the maximum kinetic energy of the outgoing 
sound pulse scales approximately as the square of the gravitational mass lost, 
as expected from equation~(\ref{eq:dEr}). The
total energy of the ejecta has an even steeper scaling 
with $\delta M_{\rm G}$.

The \emph{exponential} model is the most conservative of the three
prescriptions in terms of the amount of gravitational mass lost.
The main uncertainty in this model is the maximum mass of a cold, non-rotating 
neutron star $M_{\rm tov}$,
which is constrained from below at $2M_\odot$ by measured neutron star masses
\citep{antoniadis_2013} and at $\sim 2.5M_\odot$ from above by causality 
(e.g., \citealt{lattimer_2016}). The value adopted is at the upper range of
allowed values.   Alternative choices should lead to factors of less than two
difference in the gravitational mass lost (LW13).

\subsection{Global Ejecta Properties}

\begin{figure*}
\includegraphics*[width=\textwidth]{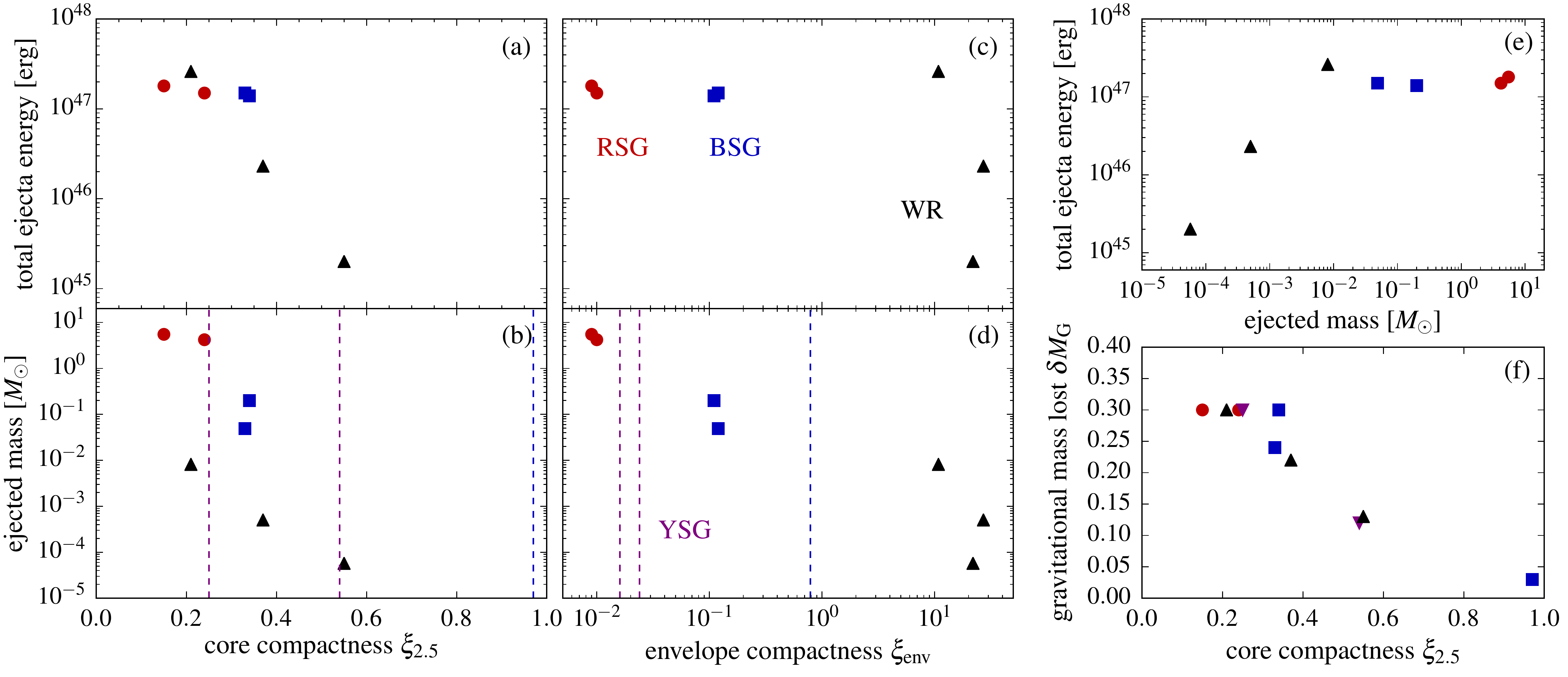}
\caption{Global ejecta properties for models that employ the \emph{exponential} neutrino
loss prescription (Table~\ref{t:models}), with red circles, blue squares, and black triangles denoting
RSGs, BSGs, and WRs, respectively. \emph{Left:} Total ejecta energy (top) and mass (bottom) as a function
of core compactness (equation~\ref{eq:compactness}). \emph{Middle}: Total ejecta energy (top)
and mass (bottom) as a function of envelope compactness (equation~\ref{eq:global_compactness}). 
\emph{Top right:} total ejecta energy versus ejecta mass. \emph{Bottom right:} gravitational mass lost as a function
of core compactness. Purple triangles denote YSGs, and vertical dashed lines correspond to failed models.}
\label{f:ener_mass_xi_dM_ejected}
\end{figure*}

The total ejecta energies and masses for all models that employ the
\emph{exponential} neutrino loss prescription are shown
in Figure~\ref{f:ener_mass_xi_dM_ejected}. Results are shown as a function of both 
core compactness $\xi_{2.5}$ and envelope compactness $\xi_{\rm env}$.

The ejecta energy is a monotonically decreasing function 
of the core compactness, and does not appear to be very sensitive to the 
envelope compactness. This result can be attributed to two physical effects. First,
the energy available to power the shock $\Delta E(r_c)$ depends on
the gravitational mass lost to neutrinos, $\delta M_{\rm G}$, 
which depends on the neutrino cooling time. Progenitors with
a high core compactness reach the TOV limit earlier, and hence
do not radiate as much energy. This is shown explicitly in Figure~\ref{f:ener_mass_xi_dM_ejected}f,
where increasing the core compactness above $0.4$ results in a nearly linear
decrease in the gravitational mass radiated in neutrinos. Below $\xi_{\rm 2.5}=0.4$,
the mass loss to neutrinos saturates at about $0.3M_\odot$. This is 
a property of the chosen \emph{exponential} prescription (equation~\ref{eq:max_loss}), 
which yields 
$\delta M_{\rm G}\simeq \textrm{BE}_c(M_{\rm tov})[1-e^{-1}]$ unless $\tau_{\rm tov}\ll \tau_c$.

Note that while a shorter neutrino cooling time in principle also decreases
$r_c$, the maximum energy $\Delta E(r_c)$ does not increase (Table~\ref{t:models}) because of
the decrease in $\delta M_{\rm G}$ and because the maximum energy released tends to
flatten out for $t_{\rm ff} < 1$~s at constant $\delta M_{\rm G}$ (Figure~\ref{fig:dE}).

The second effect that suppresses the shock energy for high compactness
is the larger gravitational energy at the pressure wave formation radius $r_c$.
Thus more kinetic and internal energies are spent climbing out of the
gravitational potential. Indeed, the only model for which the shock fails to
break through the stellar surface (B80z-4\_e) has the highest core compactness
of the set and the smallest value of $r_c$.

The final shock energies do not appear to depend sensitively on the
envelope compactness. In fact, with the exception of the WR progenitors
with higher core-compactness, the energies of successful shocks are not much
smaller than the values predicted by $\Delta E(r_c)$.

In contrast, the ejecta mass displays the opposite dependence on progenitor structure. 
A clear hierarchy is evident in Figure~\ref{f:ener_mass_xi_dM_ejected}d: 
stars that are able to eject mass (with positive energies) do so in
amounts that correlate negatively with increasing envelope compactness, 
in a manner consistent with the ordering shown in Figure~\ref{fig:bind}.
In the case of WRs, the ejected mass and energy are strongly 
correlated (Figure~\ref{f:ener_mass_xi_dM_ejected}e). The clear dependence of 
the ejected mass on core compactness for these progenitors is likely a
consequence of the dependence of the energy on core compactness.

A correlation between the total ejected energies and (unbound) mass was
predicted by \citet{Tan+01}. While we do observe this correlation for WRs and RSGs,
this is not the case for BSGs (Figure~\ref{f:ener_mass_xi_dM_ejected}f). 

\subsection{Failed Shocks}
\label{s:failed}

Out of our model sample, four cases failed to eject unbound matter by the end of 
the simulated time. Two factors can lead to failure: a high core compactness,
and/or a large envelope mass.

The first case worth noting is model Y25z-2\_e, which has about the same
core compactness as the fiducial RSG but with a larger envelope
due to its lower metallicity (envelope compactness is larger by a factor of $2.5$).
While the model ejects $2.5M_\odot$ of material, by the time the simulation
ends (and the floor of temperature in the EOS is reached) the material
is gravitationally bound (net energy $-10^{47}$~erg).

\begin{figure}
\includegraphics*[width=\columnwidth]{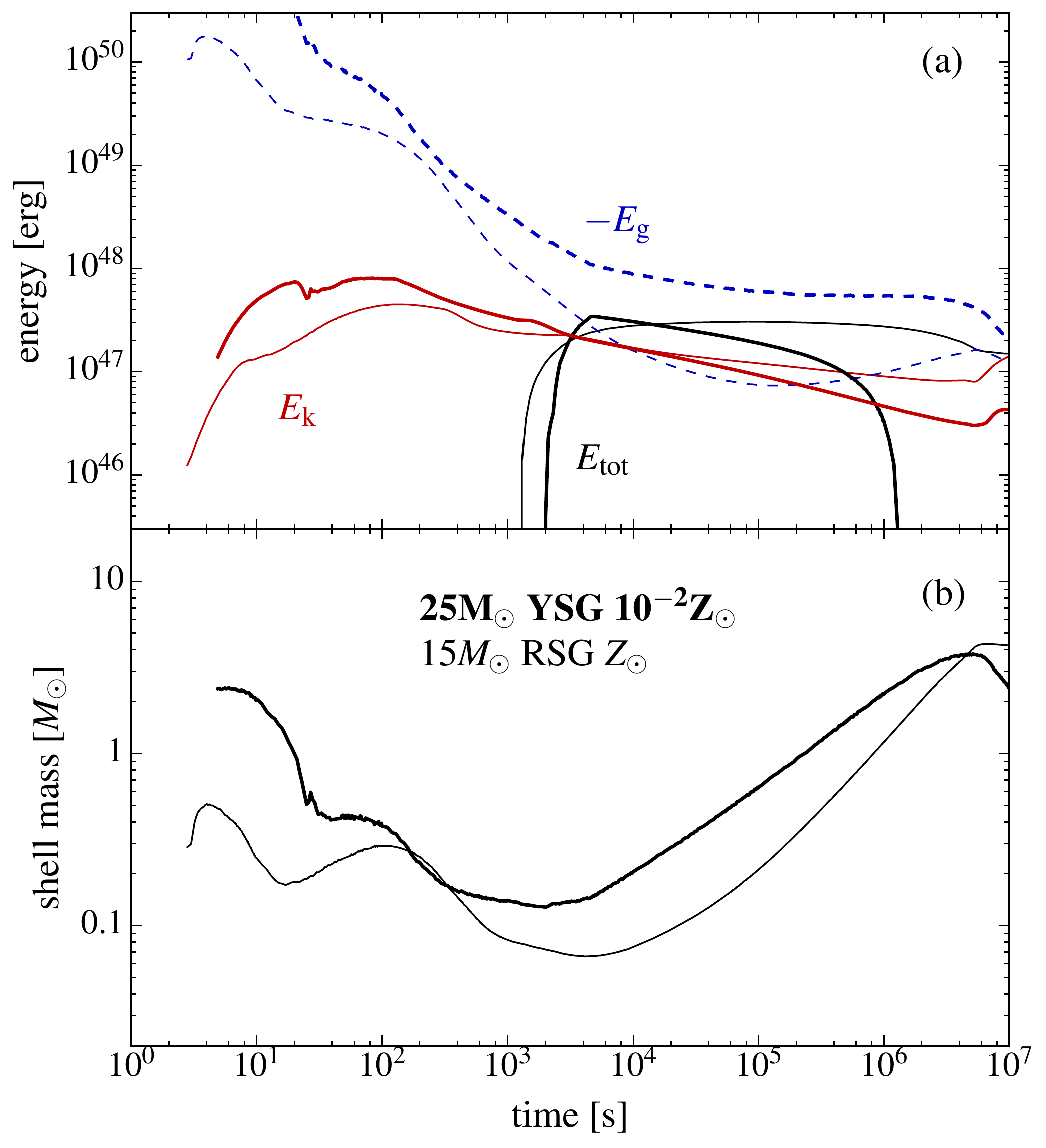}
\caption{Evolution of the energies and shock mass in models R15z00\_eHR (thin lines) and
Y25z-2\_e (thick lines), which differ primarily in their envelope compactnesses 
($\xi_{\rm env}=0.01$ and $0.024$, respectively, c.f. Table~\ref{t:progenitors}). 
\emph{Top:} Kinetic (red), gravitational (blue),
and total (black) energies in the outgoing pressure wave. \emph{Bottom:} mass in the outgoing wave.
The YSG ejecta is gravitationally bound by the time the simulation ends and the temperature floor is reached.}
\label{f:failed_m25z-4}
\end{figure}

Figure~\ref{f:failed_m25z-4} shows the energies and ejected masses of this
model in comparison with the fiducial RSG (R15z00\_eHR), also shown in 
Figure~\ref{f:ener_mass_vshock_fiducial}. The early evolution of the
two models is similar, with the outgoing shock becoming unbound inside the
star in both cases. As the shock sweeps up mass in the envelope, however,
the deeper gravitational potential in the YSG keeps the gravitational
energy in the shock high and as the kinetic energy decreases, the
shock becomes bound again before breakout.

The negative energy at breakout does not mean complete failure, as model W40z00\_eHR
shows (Figure~\ref{f:ener_mass_vshock_fiducial}). Nevertheless, the final ejected
mass can be significantly lower than that contained in the shock at the time of breakout.
Unfortunately, our simulations cannot reach the required times to determine definitively 
what happens since the  temperature floor in the Helmholtz EOS ($10^4$~K) is reached shortly
after breakout and the results become unreliable thereafter.

The other case worth noting is that of models B80z-2\_e and B80z-2\_f, which correspond to 
a low-metallicity blue supergiant with a very large core-compactness ($\xi_{\rm 2.5}=0.97$) and a 
large envelope (total mass $55M_\sun$ at core-collapse). The large core-compactness
results in a very short time to reach the TOV mass and a small amount of 
gravitational mass lost to neutrinos ($\delta M_{\rm G} = 0.03M_\odot$ for B80z-2\_e).
The resulting sound pulse has a very low energy ($E_{\rm k,max}\simeq 3\times 10^{45}$~erg for B80z-2\_e),
which is spent mostly climbing out of the large gravitational potential well.
The failure of the model is robust to changes in the evolution of the inner core: employing
the \emph{full loss} prescription, which increases the gravitational mass lost by $30\%$ 
(model B80z-2\_f) produces the same qualitative outcome.

\begin{figure}
\includegraphics*[width=\columnwidth]{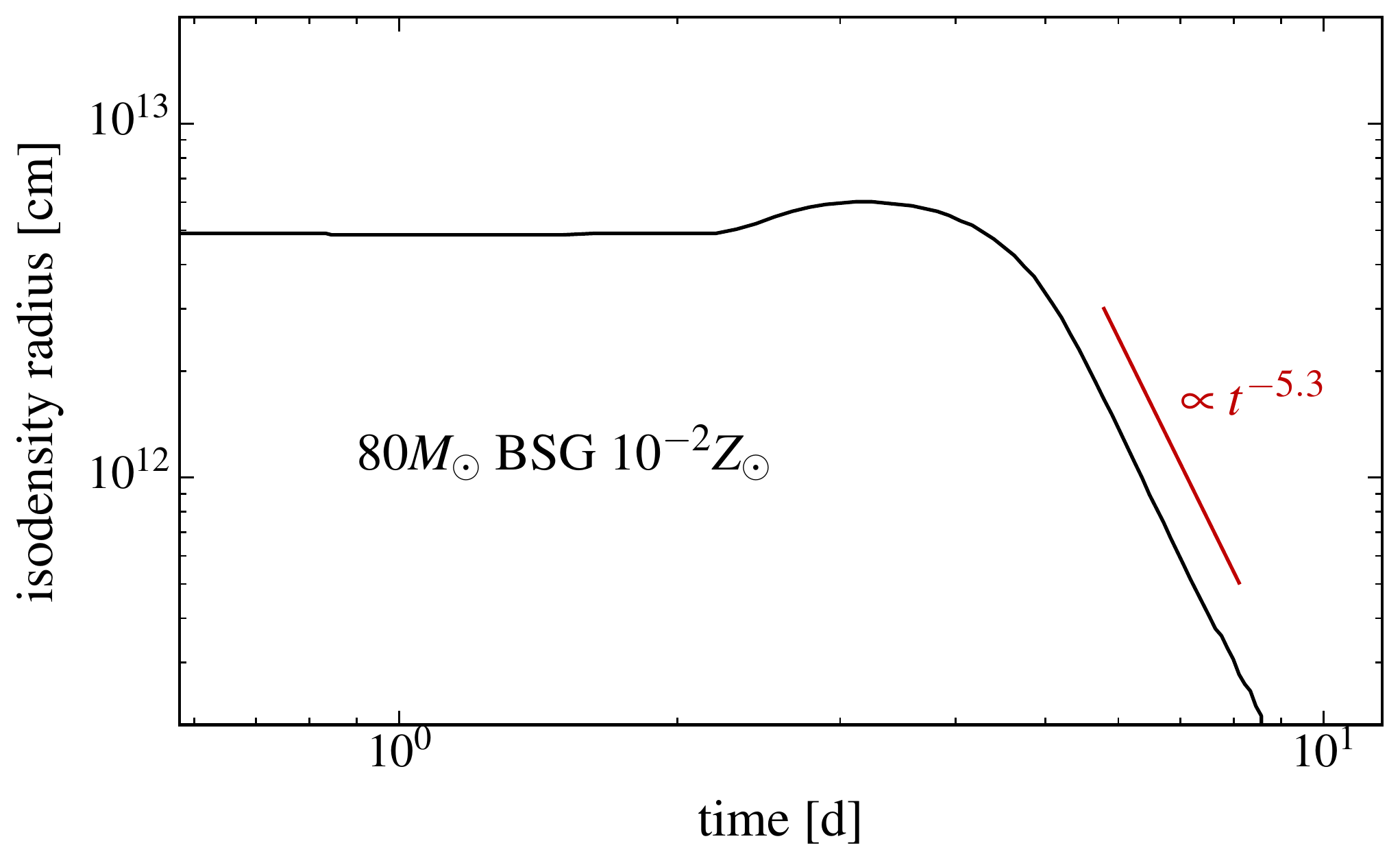}
\caption{Evolution of the stellar surface of model B80z-2\_e, which fails to eject any
mass. The stellar radius is quantified by the position of the initial photospheric density 
$\rho_{\rm ph}\simeq 2.5\times 10^{-10}$~g~cm$^{-3}$.  The small increase in the stellar radius 
is due to the weak sound pulse/shock approaching the surface. This would likely manifest itself 
as a small decrease in the stellar effective temperature prior to its disappearing.}
\label{f:failed_m80z-4}
\end{figure}

Figure~\ref{f:failed_m80z-4} shows the evolution of the stellar surface in model B80z-2\_e -- quantified as the 
position of the photospheric density -- as the shock reaches it. The low shock energy results 
in an expansion of the star by about $\sim 25\%$ in radius over a period of a few days, 
followed by a steep infall.   Such a star would likely show a modest decrease in effective temperature due to this 
expansion of the photosphere, just prior to disappearing.

The last unsuccessful case is model Y22z00\_e, which has a relatively high
core compactness ($\xi_{\rm 2.5}=0.54$) but an envelope mass similar to that
of the successful RSGs (the envelope compactness is higher by 60\% relative
to the fiducial RSG). The high core compactness also results in a short
time to reach the TOV mass and thus a small amount of gravitational mass
lost ($\delta M_{\rm G} = 0.12M_\odot$). The shock energy is therefore low.
Nevertheless, the model ejects a small amount of mass $\sim 10^{-1}M_\odot$
with marginally negative energy ($\sim 10^{45}$~erg). Since the dynamics
at late times after shock breakout for extended stars is not reliable in our simulations, 
we do not report the ejected mass and energy in Table~\ref{t:models}, in contrast to model Y25z-2\_e, 
for which mass ejection is unambiguous (though gravitationally bound). Simulations with 
a different EOS and/or numerical method will be required in order to better understand 
these marginal RSG/YSG cases, but it is clear that if there is any unbound material, it will 
have significantly less mass and energy than the majority of our other progenitors.

\subsection{Fallback Accretion}

In all of our progenitors, most of the star collapses onto the black hole.   
The resulting fallback accretion rate -- assuming no rotation -- can extend from hours 
to years, depending on the progenitor.

Given that our simulations move the inner boundary to increasingly
larger radii in order to save computing time (\S\ref{sec:hydro}), we
need to carry out an extra step in order to obtain the fallback accretion
rate close to the BH. Fortunately, the accretion rate at small radii 
depends mostly on time, and the problem is such that an excellent 
semi-analytic approximation can be obtained (Appendix~\ref{s:accretion_appendix}).

Since the radial position of the trailing edge of the shock $r_{\rm tr}$ is
defined as the innermost point with zero velocity, we only need to compute the
infall from rest of a given mass shell reached by this trailing edge in order to
obtain the fallback accretion rate. We provide a detailed derivation of this
calculation in Appendix~\ref{s:accretion_appendix}. The accretion time for the
shell is
\begin{eqnarray}
t_{\rm acc}(r,t) & = & t + t_{\rm fall}(r,r_{\rm tr}[t])\nonumber\\
& \simeq & t + \frac{r_{\rm tr}^3(t)}{\sqrt{2GM(r_{\rm tr}[t],t)}}\left[\frac{\pi}{2}
               -\frac{2}{3}\left(\frac{r}{r_{\rm tr}[t]}\right)^{3/2}\right],\nonumber\\
\label{eq:tacc_main-text}
\end{eqnarray}
where the expression is valid for $r \ll r_{\rm tr}(t)$, and the accretion rate
at time $t_{\rm acc}$ is 
\begin{equation}
\label{eq:mdot_main-text}
\dot{M}(r,t_{\rm acc}[r,t]) \simeq f_{\rm fall}\frac{16}{3}r_{\rm tr}^2(t) \rho(r_{\rm tr}[t],t)
\sqrt{\frac{2GM(r_{\rm tr}[t],t)}{r_{\rm tr}(t)}},
\end{equation}
where the use of the free-fall speed is a good approximation at late times in the 
infall ($r\ll r_{\rm tr}(t)$). The fudge factor $f_{\rm fall}$ is added to account
for the early part of the infall, in which gas pressure effects are important.

\begin{figure}
\includegraphics*[width=\columnwidth]{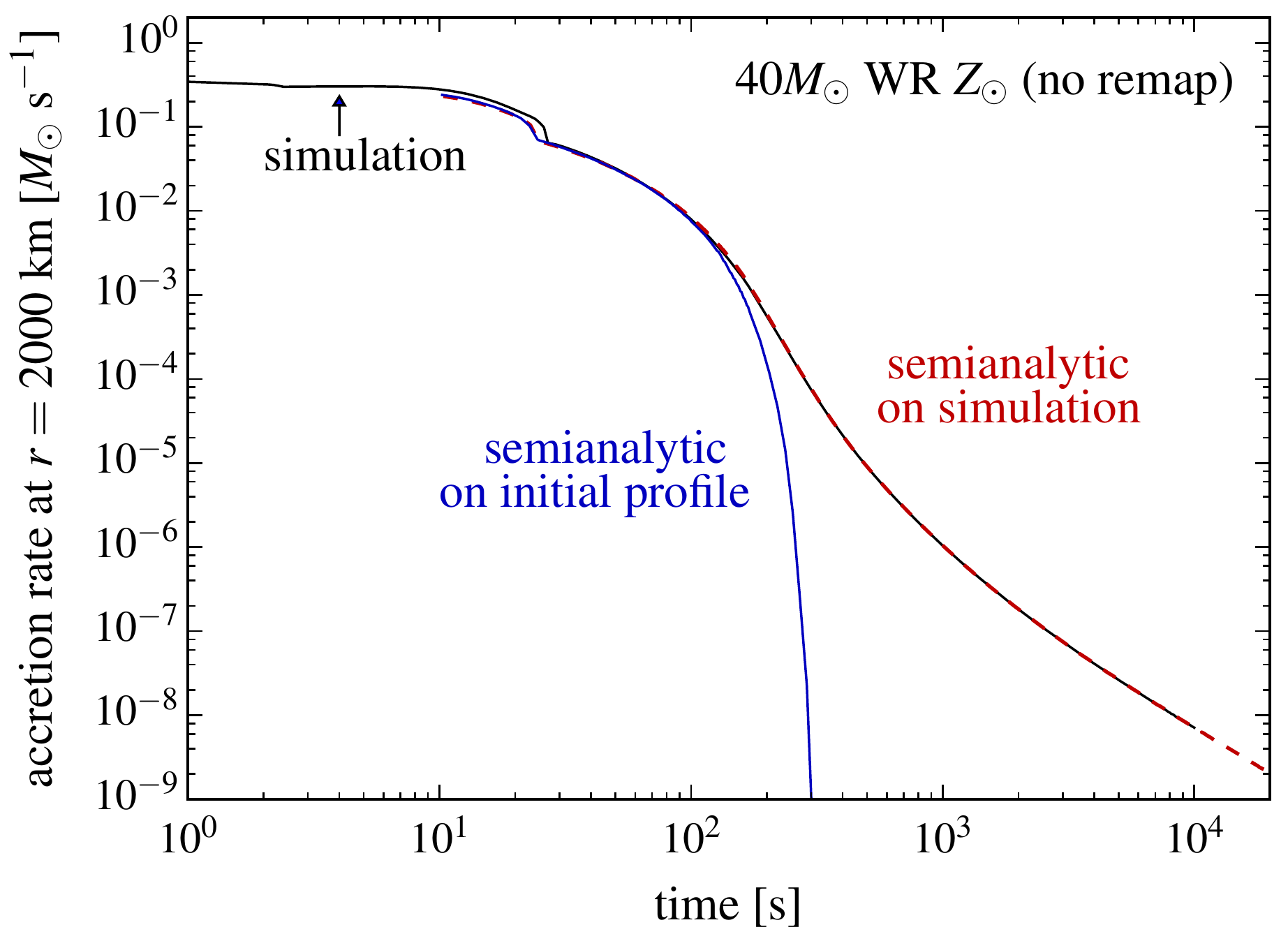}
\caption{Accretion rate -- assuming no rotation -- as a function of time in a version of model W40z00\_e for which
the inner boundary is kept constant at $r = 2000$~km. The black curve shows the accretion rate
measured in the simulation, while the red dashed curve shows the analytic approximation to
the accretion rate at small radii in equations~(\ref{eq:tacc_main-text})-(\ref{eq:mdot_main-text}) using the
density at the trailing edge of the shock and $f_{\rm fall}=1/2$. 
For comparison, the blue curve shows the result of 
applying the same equations to the initial density profile of the stellar progenitor.  
Accretion onto the newly formed black hole extends much later in time in the full simulations 
due to the marginally bound ejecta generated by neutrino-induced mass loss.}

\label{f:mdot_calibration}
\end{figure}

Figure~\ref{f:mdot_calibration} shows a quantitative test of 
equations~(\ref{eq:tacc_main-text})-(\ref{eq:mdot_main-text}). We employ a version
of model W40z00\_e for which the inner boundary is kept constant at $r=2000$~km
for the entire simulation, and compare the accretion rate measured from the 
simulation with the semianalytic approximation using $f_{\rm fall}=1/2$. Agreement is excellent 
at times $t\gg 10$~s, given that the condition $r\ll r_{\rm tr}(t)$ is well satisfied. 
Also, at times $t\lesssim 10$~s the position of the trailing edge of the shock 
(sound pulse at the time) is not so well defined. 

\begin{figure*}
\includegraphics*[width=\textwidth]{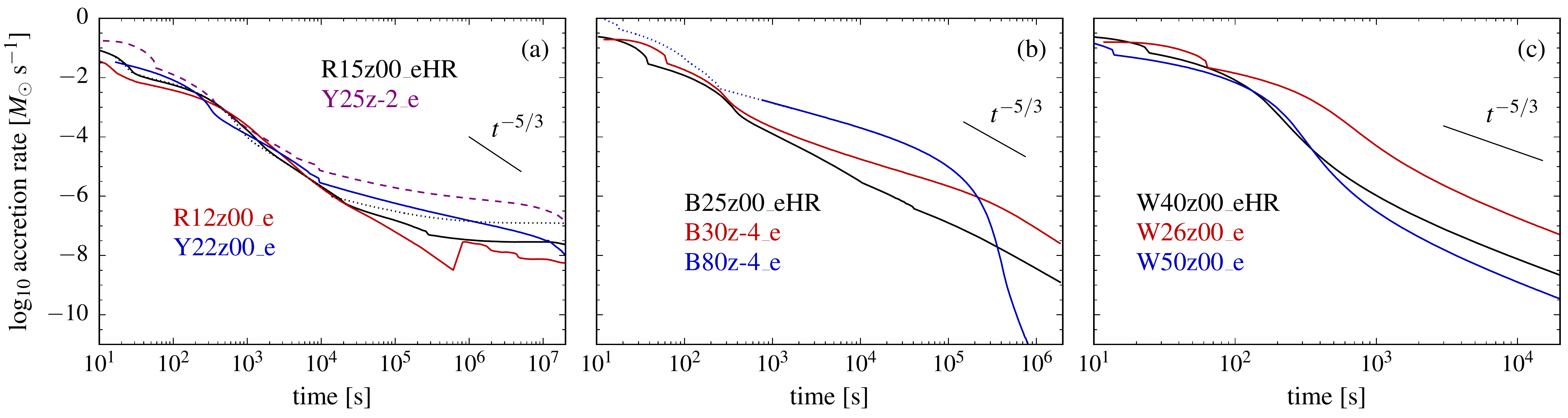}
\caption{Accretion rate at small radii -- a proxy for the black hole accretion rate assuming no rotation -- 
for various models, as labeled, obtained by using 
the semianalytic approximation in equations~(\ref{eq:tacc_main-text})-(\ref{eq:mdot_main-text}) 
(c.f.~Figure~\ref{f:mdot_calibration}) with $f_{\rm fall}=1/2$. The dotted line in the left panel is the accretion
rate obtained using the initial progenitor density profile for model R15z00\_eHR, for reference. 
The dotted line for model B80z-2\_e (middle panel) is the accretion rate measured at the inner boundary given that the trailing 
edge of the shock becomes measurable only at late times.}
\label{f:mdot_all-models}
\end{figure*}

As an experiment, we have also evaluated equations~(\ref{eq:tacc_main-text})-(\ref{eq:mdot_main-text})
using the initial density profile of the stellar progenitor. Figure~\ref{f:mdot_calibration}
shows that the resulting accretion rate drops steeply shortly after the time of shock breakout
in this model ($\sim 100$~s). This demonstrates that the enhanced accretion measured at late times comes from
material in the outgoing shell which is not gravitationally unbound and thus falls back to the 
BH (accounting for the decreasing mass in Figure~\ref{f:ener_mass_vshock_fiducial}h).

Figure~\ref{f:mdot_all-models} shows the result of applying 
equations~(\ref{eq:tacc_main-text})-(\ref{eq:mdot_main-text}) to all the models that
employ the \emph{exponential} neutrino loss scheme. Except for
the failing model B80z-2\_e, all progenitors lead to sustained accretion from hours to years. 
In the case of RSGs and YSGs, this accretion has a shallow time dependence, since the surface
of the star has not yet fallen in by the time the simulation stops (the collapse of
the initial density profile for model R15z00\_eHR is shown in Figure~\ref{f:mdot_all-models}a,
for reference). This suggests that fallback accretion will last for many years for these
extended progenitors. The fallback accretion declines more rapidly in time for BSGs and WRs,
following approximately a $t^{-5/3}$ time dependence at very late times once fallback from
the rear of the ejecta reaches the BH.

More realistically, not all of the star will fall radially onto the black hole. The specific
angular momentum required to circularize just outside of the innermost stable circular
orbit is (e.g., \citealt{margalit_2015})
\begin{equation}
\label{eq:j_isco}
j_{\rm isco} \simeq \left( 4\times 10^{16} - 10^{17}\right) M_{\rm cc,10}\,\textrm{cm}^2\textrm{ s}^{-1},
\end{equation}
where $M_{\rm cc,10} = M_{\rm cc}/(10M_\odot)$, and the numeric range accounts for the black hole spin.
Massive stars are rapid rotators, with surface rotational velocities $v_{\rm rot}\sim 100$~km~s$^{-1}$ 
on the main sequence (e.g., \citealt{fukuda_1982}). The specific angular momentum of material at the stellar surface
in a pre-supernova star is
\begin{equation}
\label{eq:j_surf}
j_{\rm surf} = 10^{18}\, v_{\rm rot,7}\, R_{\rm cc,11},\textrm{cm}^2\textrm{ s}^{-1},
\end{equation}
where $v_{\rm rot,7} = v_{\rm rot}/(10^7\textrm{ cm s}^{-1})$ 
and the pre-supernova radius is $R_{\rm cc,11} = R_{\rm cc}/(10^{11}\textrm{ cm})$.
The corresponding circularization radius of this surface material would be
\begin{equation}
r_{\rm circ} \simeq 7\times 10^{8}v_{\rm rot,7}^2 R_{\rm cc,11}^2 M_{\rm cc,10}^{-1}\textrm{ cm}.
\end{equation}

Rotation rates for WR stars are difficult to measure \citep{Crowther07,st-louis_2009} although they
are predicted to be in the range $10-100$~km~s$^{-1}$ \citep{meynet_2003} with slow rotation more likely
due to the intense mass loss. Except for the case of very slowly-rotating stars with 
$v_{\rm rot} < 10$~km~s$^{-1}$ and a non-spinning black hole, the formation of a fallback disk out of 
material in the outer layers is almost guaranteed. The larger radii of BSG progenitors and the small amount
of mass ejected makes disk formation even more likely. 

In the case of RSGs, ejection of the entire hydrogen envelope means that the last material to fall 
back is located at the base of this envelope. For our fiducial progenitor, this radius is located at 
$r_{\rm base}\sim 4R_\sun \simeq 3\times 10^{11}$~cm. Assuming uniform rotation in the envelope, the
rotational velocity at this radius is $v_{\rm rot,7} \sim (r_{\rm base}/R_{\rm cc})\sim 4\times 10^{-3}$,
which would bring $j_{\rm surf}$ (equation~\ref{eq:j_surf}) below $j_{\rm isco}$ (equation~\ref{eq:j_isco})
and an accretion disk may not form. If, on the other hand, the specific angular momentum is constant
with radius in the H envelope, disk formation is likely to occur.

The incidence of these late-time disks in failed supernovae has been considered 
previously \citep{Quataert_Kasen_2012,Woosley_Heger_2012} and estimates of the accretion lifetimes extend to 
thousands of years \citep{Perna_et_al_2014}. This fallback accretion might power long 
time-scale high energy transients.   In addition, \citet{Kashiyama_Quataert_2015}  predicted a 
UV/optical transient lasting $\sim 10$ days assuming an outflow from the radiatively inefficient 
fallback disk that circularizes at small radii.

The temporal dependence of the decay in the bolometric luminosity ($\sim t^{-4/3}$)
measured by \citet{adams_2017a} for their failed supernova candidate suggests the existence of such a fallback disk,
as this temporal slope is expected for a super-Eddington slim disk
model (e.g., \citealt{Perna_et_al_2014}).  We note, however, that the fallback rates in RSGs are likely to be super-Eddington for many years (Fig.~\ref{f:mdot_all-models}).   In this regime, the luminosity is unlikely to decline much as the accretion rate does, and may in fact be roughly constant until the fallback accretion rate is below Eddington.  It is thus somewhat puzzling that the luminosity of the \citet{adams_2017a} transient decays at a rate of order the expected fallback accretion rate on year timescales.  Future work should address the formation, long-term evolution, and emission
of the fallback accretion with multi-dimensional time-dependent simulations.

\section{Observational Implications}
\label{sec:observations}

In this section we estimate the observational manifestation of the weak
explosions calculated in the previous sections.  We focus on the most robust
predictions, which are associated with the spherically symmetric shock breakout
and recombination powered emission.   Fallback accretion might in some cases
produce a separate transient if a significant amount of mass becomes
rotationally supported at late times.   
In future work it would be interesting to quantitatively apply the fallback
accretion rates found here to such models (e.g., \citealt{Quataert_Kasen_2012,
Woosley_Heger_2012,Kashiyama_Quataert_2015}).

\begin{table*}
\centering
\begin{minipage}{15.5cm}
\caption{Bolometric emission properties inferred from the fiducial model set. Columns
from left to right show: model name, shock breakout luminosity (eq.~\ref{eq:L_bo}), 
breakout time $t_{\rm bo}=\min(t_{\rm diff},t_{\rm bo})$ (eq.~\ref{eq:tdiff_bo}-\ref{eq:tlc_bo}), 
shock velocity at breakout $v_{\rm bo}$, gas temperature at breakout $T_{\rm bo}$, 
plateau luminosity (eq.~\ref{eq:L_pl}), plateau duration (eq.~\ref{eq:t_pl}),
final shock velocity $v_{\rm exp} = \sqrt{2E_{\rm ej}/M_{\rm ej}}$, mass
fraction of hydrogen, helium, carbon, and oxygen at the stellar surface.
Shock breakout parameters are in part analytic estimates based on \citet{waxman_2016} (see \S \ref{sec:observations}), given that we are not fully resolving
the regions close to the photosphere and hence the shock acceleration in those regions.
\label{t:observations}}
\begin{tabular}{lccccccccccc}
\hline
Model & $L_{\rm bo}$                     & $t_{\rm bo}$ & $v_{\rm bo}$  & $T_{\rm bo}$ & $L_{\rm pl}$    
        & $t_{\rm pl}$ & $v_{\rm exp}$ & $X_{\rm H}$  & $X_{\rm He}$ & $X_{\rm C}$ & $X_{\rm O}$\\
      & $(L_\odot)$ &           & (km~s$^{-1}$) & (K)          & $(L_\odot)$ 
        &    (d)       & (km~s$^{-1}$) &     &     &    & \\
\hline
R15z00\_eHR & $6$E+6   & $3$d  & $70$   & $9$E+3 & 6E+5 & 400  & $70$   & 0.67 & 0.31 & 0    & 0   \\
B25z00\_eHR & $2$E+8   & $3$h  & $900$  & $7$E+4 & 2E+6 & 20   & $600$  & 0.41 & 0.57 & 0    & 0    \\
W40z00\_eHR & $3$E+8   & $1$s & $12,000$ & $1$E+6 & 5E+4 & 2    & $2000$ & 0    & 0.18 & 0.49 & 0.30 \\   
\hline
\end{tabular}
\end{minipage}
\end{table*}

\subsection{Shock Breakout}

The emergence of a successful shock from the stellar surface is accompanied
by a brief burst of radiation once photons trapped at the leading
edge of the shock diffuse out \citep{colgate_1974,falk_1978}. Radiation starts escaping once the shock reaches
an optical depth $\tau_{\rm bo} = c/v_s$ from the stellar surface (e.g., \citealt{sapir_2011}). The
duration of this signal is the longest of the diffusion time over the
characteristic width of a radiation-dominated shock $\delta \tau = \tau_{\rm bo}$ \citep{weaver_1976}
\begin{equation}
\label{eq:tdiff_bo}
t_{\rm diff,bo} \simeq \tau_{\rm bo}\frac{(R_{\rm cc}-R_{\rm bo})}{c}
\end{equation}
and the light-crossing time over the stellar surface \citep{ensman_1992}
\begin{equation}
\label{eq:tlc_bo}
t_{\rm lc} \simeq \frac{R_{\rm cc}}{c},
\end{equation}
with $R_{\rm bo}$ the radius where the optical depth is $\tau_{\rm bo}$.
The breakout luminosity is simply the radiation energy within the transition
region $E_{\rm rad, bo}$ divided by the breakout time (e.g., \citealt{Piro_2013})
\begin{equation}
\label{eq:L_bo}
L_{\rm bo}\simeq \frac{E_{\rm rad, bo}}{\max\{t_{\rm diff}, t_{\rm lc}\}}.
\end{equation}

Table~\ref{t:observations} shows estimates for the bolometric shock breakout luminosity
and timescale for the baseline model set. Given that we are not fully resolving the
layers near the photosphere (\S\ref{s:overview}), we estimate the shock breakout velocity $v_{\rm bo}$
with the formulae in \citet{waxman_2016} using the ejecta energy and mass from Table~\ref{t:models}
and the progenitor radius $R_{\rm cc}$.\footnote{We also assume $f_\rho$ = $\kappa_{0.34}=1$ in their equations.}
The resulting values are in good agreement with the value
measured in the simulation for the RSG and within a factor of two for the BSG. In the
case of the WR, for which we have the poorest resolution at the photosphere, the analytic
estimate is a factor $\sim 4$ larger than the velocity in the simulation.

Given the breakout velocity, the resulting
optical depth $\tau_{\rm bo} = c/v_{\rm bo}$ is used to measure the radial distance from the 
surface of the star at which breakout occurs. This yields the diffusion time and allows measuring
the radiation energy $E_{\rm bo}$ contained in the transition region for
use in equation~(\ref{eq:L_bo}). For the WR progenitor, this procedure results in only
one cell in the simulation contributing to the radiation energy, which we consider
unreliable, so for this progenitor we use instead the analytic estimate from \citet{waxman_2016}.
The breakout temperature is estimated assuming
black body radiation $L_{\rm bo} = 4\pi R_{\rm cc}^2\, \sigma T_{\rm bo}^4$. 

Shock breakout in the RSG can reach peak luminosities $\sim 2\times 10^{40}$~erg~s$^{-1}$
($10^6-10^7L_\odot$)
and last for a few days. These values agree favorably with the
estimates from \citet{Piro_2013} and \citet{lovegrove_2017} given the mass and 
energy of the ejecta in our models.

Shock breakout in the BSG model produces a luminosity $\sim 10^{42}$~erg~s$^{-1}$ ($10^8L_\odot$)
over a timescale of $\sim 3$~h, presumably in the UV given $T_{\rm bo}\sim 7\times 10^4$~K. 
Our models do not include radiation diffusion and hence the detailed values for the 
temperature might change when more physics is included. Nonetheless, a transient with this
brightness and timescale can be a promising target for future wide-field, very short-cadence
surveys (e.g., \citealt{sako_2016}).

The WR model reaches a similar bolometric breakout luminosity as the BSG
($\sim 10^{42}$~erg~s$^{-1}$ or $10^{8}L_\odot$) over $\sim 1$~s. For this model
the light crossing time sets the light curve timescale. Given the estimated temperature, the emission 
should come out in the UV to soft X-rays, although detailed calculations of  radiation mediated shocks are needed to make firm predictions. Note that these estimates assume 
that the circumstellar medium is a vacuum. In reality, the star will be surrounded by material 
ejected by the powerful stellar winds that occurred earlier in its life. The properties of 
shock breakout in a dense wind can be different, including acceleration 
of particles to high energies in collisionless shocks (e.g., \citealt{katz_2012}).

\subsection{Plateau Emission}

As the ejecta expands, it converts part of its internal energy into kinetic energy and
radiates the rest away. Emission occurs 
from a photosphere at the recombination front of its dominant species, 
above which the opacity drops sharply \citep{grassberg_1971}. This results in plateau emission, 
with a luminosity and timescale
\citep{popov_1993,kasen_2009,Kleiser_Kasen_2014} 
\begin{eqnarray}
\label{eq:L_pl}
L_{\rm pl}& \simeq  & 1.8 \times 10^{39}\, E_{\rm ej,47}^{5/6}\,M_{\rm ej, 1}^{-1/2}\,R_{\rm cc,500}^{2/3}\,
   \kappa_{0.4}^{-1/3}\,T_{6000}^{4/3}\textrm{ erg s}^{-1}\nonumber\\
 & & \\
\label{eq:t_pl}
t_{\rm pl} & \simeq & 220 \, E_{\rm ej, 47}^{-1/6}\,M_{\rm ej, 1}^{1/2}\,R_{\rm cc,500}^{1/6}\,
   \kappa_{0.4}^{1/6}\,T_{6000}^{2/3}\textrm{ days}.
\end{eqnarray}
where $E_{\rm ej, 47} = E_{\rm ej}/(10^{47}\textrm{ erg})$, $M_{\rm ej,1}=M_{\rm ej}/(1 \, M_\odot)$,
$R_{\rm cc,500} = R_{\rm cc}/(500R_\odot)$, and $\kappa_{\rm 0.4}$ is the opacity
in units of $0.4$~cm$^2$~g$^{-1}$. Equations~(\ref{eq:L_pl})-(\ref{eq:t_pl}) 
account for the diffusion of radiation in an expanding medium with a
receding photosphere, and assume that the ejecta are radiation pressure dominated,
which remains true for our models despite the low  ejecta energies (although only 
marginally for the RSG case). The luminosity in equation~(\ref{eq:L_pl}) is a black body 
at the recombination surface with the recombination temperature $T_{\rm 6000}\times 6000$~K. 
For hydrogen and oxygen-dominated ejecta, this recombination temperature is approximately 
$6000$~K, while for helium-dominated ejecta it increases to $10^4$~K (e.g., \citealt{Kleiser_Kasen_2014}). The absence of 
radioactive energy injection results in a sharp drop in the emission once the recombination
front reaches the base of the ejecta (e.g., \citealt{kasen_2009,piro_nakar_2013}).

Table~\ref{t:observations} shows the bolometric luminosities and timescales associated 
with plateau emission for the three baseline progenitors. A recombination temperature
of $6000$~K is used in the RSG and WR models, while $10,000$~K is used for the BSG
model given its higher surface abundance of helium. Also shown is the final velocity of the
ejecta, which is assumed to be $v_{\rm exp} = \sqrt{2E_{\rm ej}/M_{\rm ej}}$.

Plateau emission for the RSG model is again consistent with the results of
LW13 and the estimates of \citet{Piro_2013}, with a duration of about $400$ days
and a luminosity $\sim 2\times 10^{39}$~erg~s$^{-1}$, brighter by a factor $\sim 4$ 
relative to the progenitor star. The inferred final
velocity ($\lesssim 100$~km~s$^{-1}$) is very low compared to normal supernovae.

The BSG progenitor has a plateau that can last for about $20$ days, reaching
a luminosity of $\sim 10^{40}$~erg~s$^{-1}$, brighter by a factor $\sim 5$
relative to the stellar progenitor. Such a brightening might be detectable
if the star is monitored every few days.

Finally, the WR progenitor has a plateau phase lasting for less than one day,
and with a luminosity that is about $10$ times \emph{fainter} than the
progenitor. In this case, we expect a spike of radiation following shock breakout,
followed by a steep decrease of the luminosity to a plateau a few magnitudes
fainter than the progenitor. After a day, the star should disappear.

\subsection{Failed shocks}

The case of failed shocks might still be interesting observationally.
As shown in Figure~\ref{f:failed_m80z-4}, arrival of the pressure wave
to the stellar surface results in an expansion of the star. For this
particular model, the radius expands by $25\%$ over a timescale of
days. The luminosity is likely to be unchanged due to rapid photon diffusion, 
so that the increase in surface area will lead to a modest decrease in effective temperature prior to the star disappearing.  


\section{Summary and Discussion}\label{sec:sum+dis}

We have studied the properties of the ejecta generated by
non-rotating massive stars that undergo core-collapse and
fail to produce a successful supernova.  Neutrino radiation
during the protoneutron star phase decreases  the 
mass of the core of the star by $\sim 0.1-0.5 \, M_\odot$ over a few seconds.   
The part of the progenitor exterior to a radius $\sim$ few $10^9$ cm experiences this change
as an effectively instantaneous decrease in the mass of the star.  These layers of the star are thus 
over-pressured, resulting in an outward going sound pulse that steepens 
into a shock as it travels out through the star. We have used time-dependent
hydrodynamic simulations that follow the propagation
of the outgoing pressure wave through the entire star, using an approximate prescription 
for the neutrino radiation from the inner protoneutron star. Our analysis extends 
the earlier work of LW13 by studying this mechanism of mass ejection in failed supernovae 
for a wide range of stellar progenitors. We also provide a more detailed physical 
understanding of, and analytic estimates for, the mass ejection process.  
Our main results are the following:
\newline

\noindent 
1. Successful mass ejection due to the loss of gravitational mass to neutrinos
can occur in all types of stellar progenitors, not just red supergiants
(Figure~\ref{f:ener_mass_vshock_fiducial} and Table~\ref{t:models}).
\newline

\noindent
2.  The explosion energy is a monotonically decreasing
      function of the core compactness, and the ejected mass
      is a monotonically decreasing function of the envelope  
      compactness (or equivalently, of the escape speed at the stellar surface; 
      Figure~\ref{f:ener_mass_xi_dM_ejected}).
      \newline

\noindent
3.  The maximum kinetic energy of the shock is set by the
      change in the gravitational acceleration over a free-fall
      time, at a radius where the free-fall time equals the
      neutrino cooling time (Figures~\ref{f:shock_formation_diagram} and \ref{fig:dE}). 
      This is $\sim 10^{47}-10^{48}$ erg for most progenitors.   Propagation 
      through the stellar envelope decreases the kinetic energy from
      its maximum as the pressure wave (and later shock) moves out in the gravitational potential
      (Figure~\ref{f:ener_mass_vshock_fiducial}); hence the 
      analytic estimate (equation~\ref{eq:dEr}) is an upper limit on the final ejecta energy. 
      This in turn translates into an upper limit on the ejected mass, which is 
      $\sim 5$, $0.2$, and $0.01 \, M_\odot$ for RSGs, BSGs, and WR stars, 
      respectively (Figure~\ref{fig:bind}).     
      \newline
      
\noindent
	4. For RSGs, the change in gravitational mass due to neutrino radiation unbinds
	the hydrogen envelope, which likely has the vast majority of the angular
	momentum of the progenitor.  In this case, it is likely that the resulting
	black hole will be relatively slowly spinning.   For BSGs and WRs, however, the
	ejected mass and angular momentum are negligible, so that the resulting black
	hole mass and spin is very close to that implied by the total mass and angular
	momentum of the pre-collapse progenitor (with the caveat that if the stellar
	angular momentum implies  a black hole dimensionless spin $\gtrsim 1$ this
	mapping cannot hold).   These conclusions are important for interpreting
	gravitational wave and X-ray binary inferred black hole masses and spins. 
       \newline

\noindent
	5. Stars that have a high core compactness or high envelope
	compactness relative to the average of its class 
        ($\xi_{\rm 2.5}\gtrsim 0.5$, $\xi_{\rm env}\gtrsim 0.02$ for RSGs; $\xi_{\rm 2.5}\gtrsim 0.5$,
         $\xi_{\rm env}\gtrsim 0.5$ for BSGs) fail to eject unbound matter or any
	matter at all (e.g., Figure~\ref{f:failed_m80z-4}). While none of our WRs fail, they
        all eject relatively small amounts of unbound mass.
      \newline

\noindent
6.  Successful mass ejection also results in fallback accretion over
      periods of time ranging from hours to years (Figure~\ref{f:mdot_all-models}). 
     Depending on the uncertain angular momentum distribution of the stellar progenitor, this fallback accretion might power a variety of transients, including ultra-long-duration gamma-ray bursts and 
      rapid optical transients \citep{Quataert_Kasen_2012,Woosley_Heger_2012,Kashiyama_Quataert_2015}.
      \newline

\noindent
	7.  We estimate the shock breakout and recombination-powered plateau emission
	for our fiducial RSG, BSG, and WR progenitors (Table \ref{t:observations}).
	These are the most robust observational signatures of failed supernovae. For
	RSGs our estimates are in good agreement with the previous work of LW13,
	\citet{Piro_2013}, \& \citet{lovegrove_2017}.  We find that BSGs have shock
	breakouts that last for hours with luminosities comparable to those of normal
	supernovae. The plateau emission is  a factor of several brighter than the
	progenitor star, lasting for several weeks.   In the case of WRs, shock
	breakout is extremely bright but very short-lived and likely at UV to soft X-ray energies. The plateau emission
	is bolometrically a factor $\sim 10$ {\em fainter} than the progenitor star, and
	lasts for about a day, after which the star would likely truly disappear.  An
	interesting possibility to explore in future work is the interaction between
	the weak explosions found here and the pre-collapse stellar wind, particularly
	for WR stars.  This circumstellar interaction might well be brighter than the
	plateau emission estimated here.
        \newline

Our predictions can be improved in several ways. The simplest is to update
the equation of state to include neutral hydrogen, allowing us to follow shocks
from red supergiant and blue supergiant stars until they reach homologous
expansion.  Similarly, inclusion of radiation diffusion with suitable opacities
would allow a more accurate evolution of the internal energy of the ejecta
after shock breakout.  Finally, use of a proper core-collapse supernova code to
calculate self-consistently the loss of gravitational mass would remove the
uncertainty in the change in gravitational mass of the stellar core $\delta
M_{\rm G}$. The  final ejecta energy and ejecta mass are sensitive to the exact
value of $\delta M_{\rm G}$.

\section*{Acknowledgements}
We thank Scott Adams  and Stephen Ro for useful conversations, Chris Kochanek for comments on the manuscript,
and Evan O'Connor for pointing out an error in the evaluation of equation~(\ref{eq:dEr}).
We also thank the anonymous referee for helpful comments on the manuscript, including
corrections to our derivation of the fallback accretion rate (Appendix~\ref{s:accretion_appendix}).  
RF acknowledges support from NSERC of Canada and from the Faculty of Science
at the University of Alberta. EQ was supported in part by a Simons Investigator 
award from the Simons Foundation, and
the David and Lucile Packard Foundation.  This work was also supported in part by 
the Gordon and Betty Moore Foundation through Grant GBMF5076. 
KK acknowledges support from the Japanese Society for the Promotion of Science (JSPS)
KAKENHI Grant-in-Aid for Scientific Research (No. JP17K14248).
ERC was supported by NASA through the Einstein Fellowship Program, grant
PF6-170150.
We thank the Institute for Nuclear Theory at the University of Washington for 
its hospitality and the Department of Energy for partial support during the 
completion of this work. We acknowledge stimulating workshops at Sky House and 
Oak Creek Ranch where some of these ideas germinated. The software used in this work was in
part developed by the DOE NNSA-ASC OASCR Flash Center at the University of
Chicago. 
This research was enabled in part by support provided by WestGrid (www.westgrid.ca) 
and Compute Canada (www.computecanada.ca).
This research also used resources of the National Energy Research
Scientific Computing Center (NERSC), which is supported by the Office of
Science of the U.S. Department of Energy under Contract No. DE-AC02-05CH11231;
initial computations were performed at \emph{Carver} and \emph{Edison} (repository m2058).  
This research also used the \emph{Savio} computational cluster resource provided by
the Berkeley Research Computing program at the University of California,
Berkeley (supported by the UC Berkeley Chancellor, Vice Chancellor of Research,
and Office of the CIO).

\appendix

\section{Numerical Tests}
\label{s:appendix_numerical}

\subsection{Neutrino mass loss and choice of inner radial boundary}

\begin{figure}
\includegraphics*[width=\columnwidth]{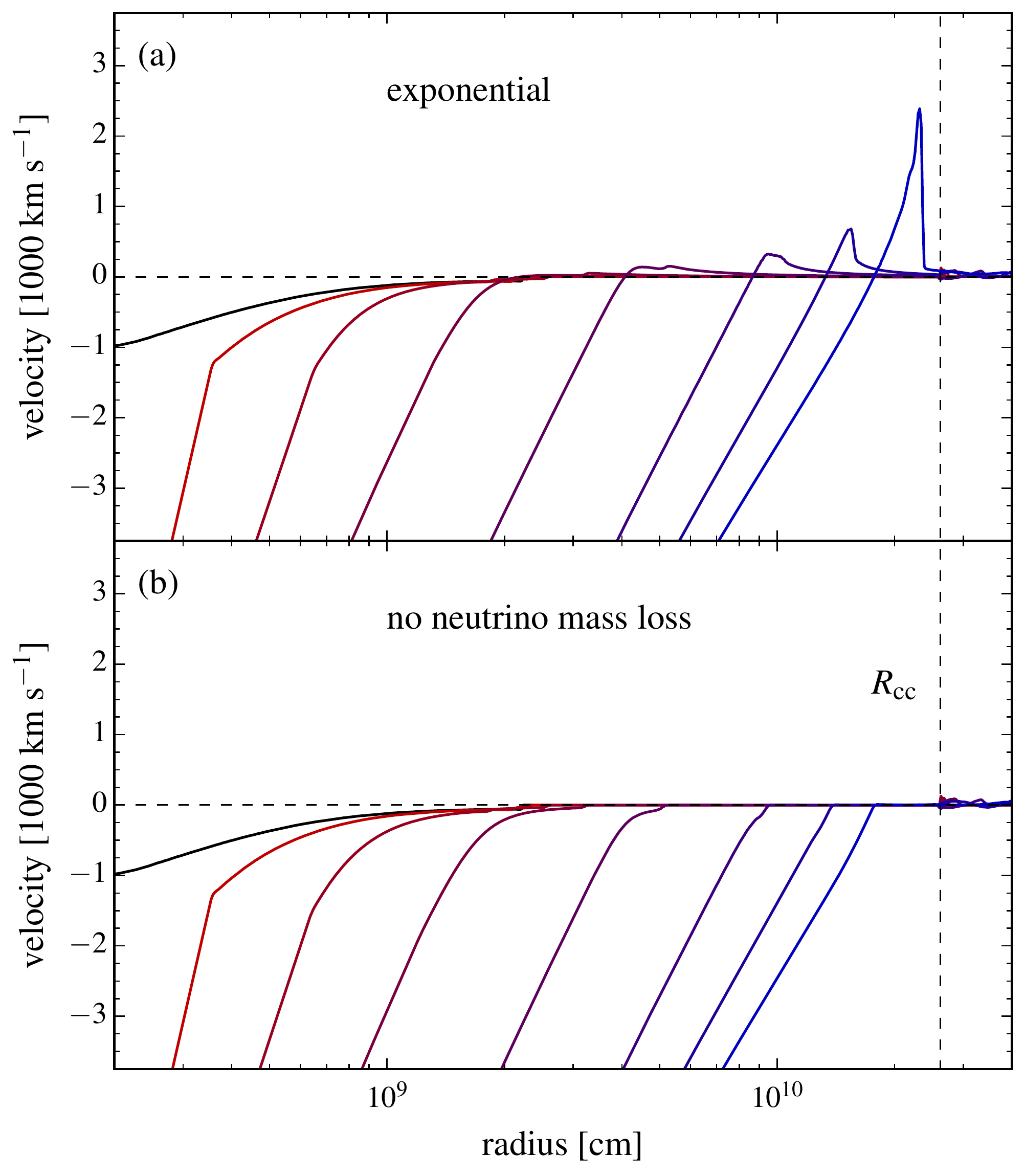}
\caption{Evolution of the velocity profile for progenitor M40z00 with neutrino
mass loss (equation~\ref{eq:max_loss}; top panel) and without ($M_{\rm G} = M_{\rm B}$; bottom panel). 
The initial progenitor profile is shown in black, and colored curves from
red to blue correspond to times 0.3, 1, 3, 10, 30, 60, and 100~s. The vertical dashed
line marks the stellar surface ($r=R_{\rm cc}$).}
\label{f:velx_nocool_appendix}
\end{figure}

We first test that no shock is generated without loss of gravitational binding energy to neutrinos.
Figure~\ref{f:velx_nocool_appendix} shows the
result of turning off the neutrino mass loss for the collapse of the baseline WR progenitor
(M40z00). In contrast to the model in which the gravitational mass decreases,
a model with no neutrino mass loss does not develop a shock. Positive velocities
outside the rarefaction wave have amplitudes smaller than $10$~km~s$^{-1}$ at $t=100$~s.
Larger velocities develop outside the star in the low-density ambient medium, but this has 
no significant effect on the dynamics.

\begin{figure}
\includegraphics*[width=\columnwidth]{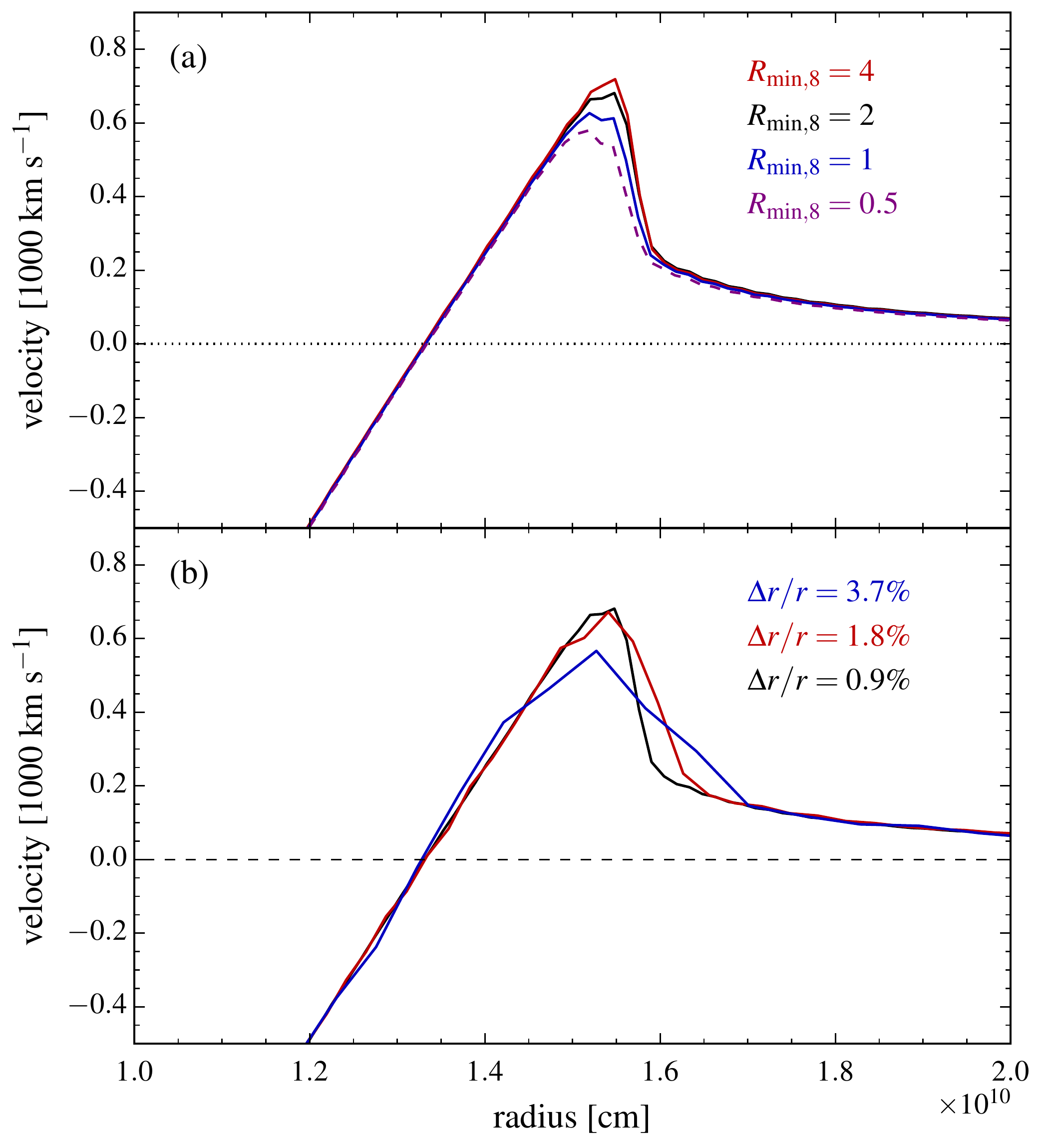}
\caption{Radial velocity as a function of radius centered around the outgoing
shock at time $t=60$~s in the baseline WR model (M40z00), using the \emph{exponential} loss prescription
(equation~\ref{eq:max_loss}). \emph{Top:} Curves show different values of the inner radial boundary, as
labeled ($R_{\rm min,8} = R_{\rm min}/10^8$~cm). \emph{Bottom:} Curves
show different spatial resolutions, as labeled.}
\label{f:velx_rmin_appendix}
\end{figure}

When neutrino cooling is included, the position of the inner boundary can influence 
the results because it determines the initial value of $M_{\rm G}$ and $M_{\rm B}$, and
thus it influences the amount of mass lost to neutrinos, all else being
equal. Following
LW13, we adopt $R_{\rm min} = 2\times 10^8$~cm, which is very near the outer
edge of the iron core for most stars. Figure~\ref{f:velx_rmin_appendix}a
shows the effect of changing the position of this boundary radius
for the baseline WR progenitor using the \emph{exponential} loss model (equation~\ref{eq:max_loss}). 
At $t=60$~s, the amplitude of the
shock increases monotonically with increasing inner boundary radius.
The maximum shock amplitude at this time lies in the range $580-720$~km~s$^{-1}$ 
for $R_{\rm min}=500$~km to $4000$~km, respectively, with the corresponding kinetic 
energies in the range $(2.5-3.7)\times 10^{47}$~erg. 

For the three smaller values of $R_{\rm min}$ shown in Figure~\ref{f:velx_rmin_appendix}a, 
the evolution of $M_{\rm G}(t)$ becomes nearly identical
after $t=1$~s, with the time to reach the TOV mass being nearly the same ($2.6-2.7$~s),
since accretion inside $r=2\times 10^8$~cm occurs faster than outside this radius.
A larger initial value of $M_{\rm B}$ (larger $R_{\rm min}$) results in a larger amount 
of mass lost to neutrinos in the \emph{exponential} loss model (equation~\ref{eq:max_loss}), 
and a larger decrease in the gravitational acceleration. 
The model with $R_{\rm min}=4\times 10^8$~cm starts out with a larger baryonic mass,
but its rate of increase is similar to that of the model with $R_{\rm min}=2\times 10^8$~cm,
hence it reaches the TOV limit at a significantly earlier time ($2.1$~s).
Despite losing less mass to neutrinos than the model with $R_{\rm min}=2\times 10^8$~cm ($0.216 M_\odot$ 
versus $0.222 M_\odot$, respectively), the shorter timescale over which the mass changes
results in a larger energy release (kinetic energy $3.7$ vs $3.4\times 10^{47}$~erg; \S \ref{sec:energy_scale}).

We conclude that the position of the inner boundary can introduce an uncertainty of $\sim 10\%$
in the outflow energy. Improving upon this uncertainty requires treating neutrino mass loss with
full physics simulations.

\subsection{Resolution and baryonic mass conservation}

To diagnose the degree of convergence of our results with
spatial resolution, we compute the same baseline WR model
as in the previous subsection  using $64$, $128$, and $256$ cells
per decade in radius, corresponding to a fractional cell size
$\Delta r/r = 3.7\%$, $1.8\%$, and $0.9\%$, respectively (our baseline
resolution is the latter). Figure~\ref{f:velx_rmin_appendix}b shows
the velocity profile around the shock at time $t=60$~s, with higher
resolution resulting in a shock with larger amplitude but narrower
width. The kinetic energies of material with positive velocity at $t=60$~s 
for the three spatial resolutions from low to high are respectively $2.9$, $3.6$, 
and $3.4\times 10^{47}$~erg. The kinetic energy is higher in the model 
with $\Delta r/r = 1.8\%$ because the mass with positive velocity 
is larger than in the model with highest resolution, $0.224$ vs $0.212 M_\odot$, 
respectively. This is visible in Figure~\ref{f:velx_rmin_appendix}b in a 
wider shock for $\Delta r/r = 1.8\%$. We therefore conclude that 
at our base resolution, results are converged to within $10\%$ in shock 
kinetic energies. 

Increasing the spatial resolution also decreases the magnitude of the
velocity fluctuations outside the star (visible in Figure~\ref{f:velx_nocool_appendix})
and the degree to which the outer stellar boundary moves before being reached
by the shock. Since these regions have very low densities, 
we do not optimize our resolution to minimize these transients, as they do not
affect the overall energetics of the shock. The velocity of the forward shock
at breakout depends on how the very surface layers of the star are resolved 
(c.f.\ref{s:overview}); we defer a more detailed study of this process for
future work and focus on the global energetics of the shock, which are 
captured with our current resolution to within $\sim 10\%$.

\begin{figure}
\includegraphics*[width=\columnwidth]{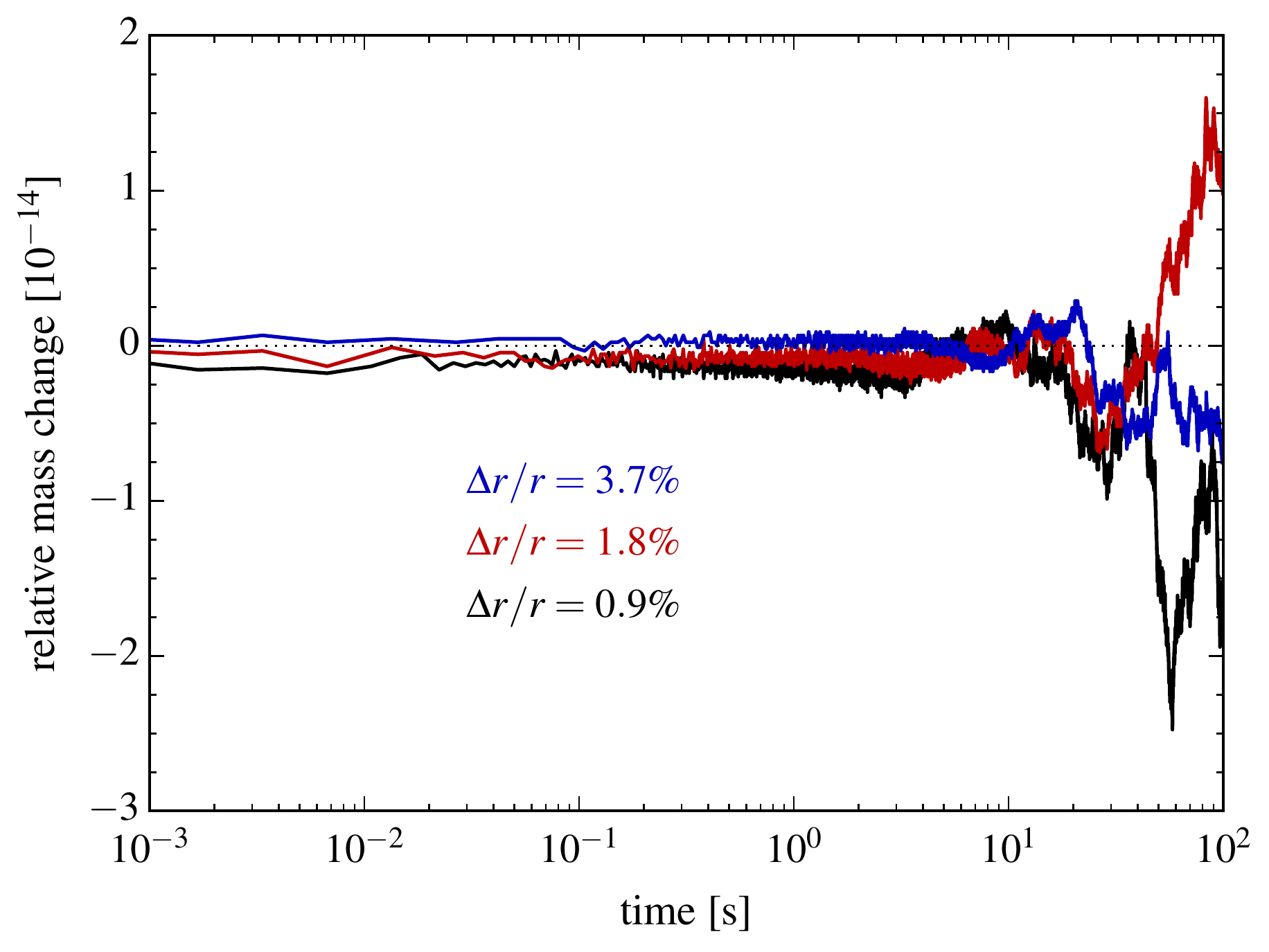}
\caption{Mass conservation as a function of time, for different spatial resolutions, 
as labeled. Curves show the relative change in the sum of the total mass in the 
computational domain plus the baryonic mass inside the inner boundary 
$M_{\rm B}$ (equation~\ref{eq:mass_conservation_diagnostic}), using the baseline 
WR progenitor and turning off neutrino mass loss. The outer 
boundary is set to reflecting, hence deviations from zero are due to 
accumulated error in integrating equation~(\ref{eq:mb_dot}). For the time period shown, the
total number of time steps are $\sim 3\times 10^4$, $6\times 10^4$, and $1.2\times 10^5$ for 
$\Delta r /r = 3.7\%$, $1.8\%$, and $0.9\%$, respectively.}
\label{f:mass_conservation_resolution}
\end{figure}

Finally, the spatial resolution has a very moderate effect on the overall
mass conservation in our simulation. Given that we employ the mass
flux from the split PPM Riemann solver in FLASH to update the point
mass for accretion (eq.~[\ref{eq:mb_dot}]), mass conservation should
in principle hold to machine precision. Figure~\ref{f:mass_conservation_resolution}
shows the quantity
\begin{equation}
\label{eq:mass_conservation_diagnostic}
\frac{M_{\rm dom}(t)+M_{\rm B}(t)}{M_{\rm dom}(0)+M_{\rm B}(0)}-1,
\end{equation}
where $M_{\rm dom}(t)$ is the baryonic mass in the computational domain,
for a baseline WR model with no neutrino mass loss and for which the outer
boundary has been set to reflecting. Any deviations from zero are thus
accumulated errors in the integration of equation~(\ref{eq:mb_dot}). Conservation
to near machine precision is indeed maintained up to about $t=10$~s, after which
the large number of time steps ($10^5$ for the highest resolution) results in oscillations. 
The spatial resolution does not appear to have a significant effect on the degree to which mass
is conserved, since all fluctuations are kept smaller than $10^{-13}$. Since
other uncertainties cause much greater changes in our results, we consider
this effect to be a negligible source of error.

\subsection{Remapping}

Following the shock evolution all the way to the surface of a RSG
is a non-trivial calculation. At the typical shock velocities obtained at large radii
($\sim 10^7$~cm~s$^{-1}$) and for characteristic RSG sizes ($\sim 10^{14}$~cm; Table~\ref{t:models})
the required physical evolution time is $\sim 10^7$~s. At our baseline resolution, the Courant
time step is approximately
\begin{equation}
\label{eq:cfl_time_step}
\Delta t_{\rm CFL}\sim 10^{-3}\left(\frac{10^9\textrm{cm s}^{-1}}{v[R_{\rm min}]}\right)
                   \left(\frac{R_{\rm min}}{2000\textrm{ km}}\right)\left(\frac{\Delta r/r}{1\%}\right)\textrm{ s},
\end{equation}
which means that $\sim 10^{10}$ time steps would be needed for the shock to reach the stellar surface
if the size of the computational domain was kept fixed.

The problem simplifies due to the straightforward collapse of the inner 
layers of the star toward the BH, quickly reaching supersonic velocities.
This means that these inner layers become causally disconnected from the rest
of the star. The acceleration of gravity at any point depends only
on the enclosed gravitational mass and not on the detailed mass distribution
as long as spherical symmetry is maintained. One can therefore move the position
of the inner boundary outward as long as infall is supersonic, allowing
longer time steps (equation~\ref{eq:cfl_time_step}), without affecting
the dynamics. A similar approach was used by \citet{hammer_2010} for evolving
a successful core-collapse supernova shock over long timescales.

\begin{figure}
\includegraphics*[width=\columnwidth]{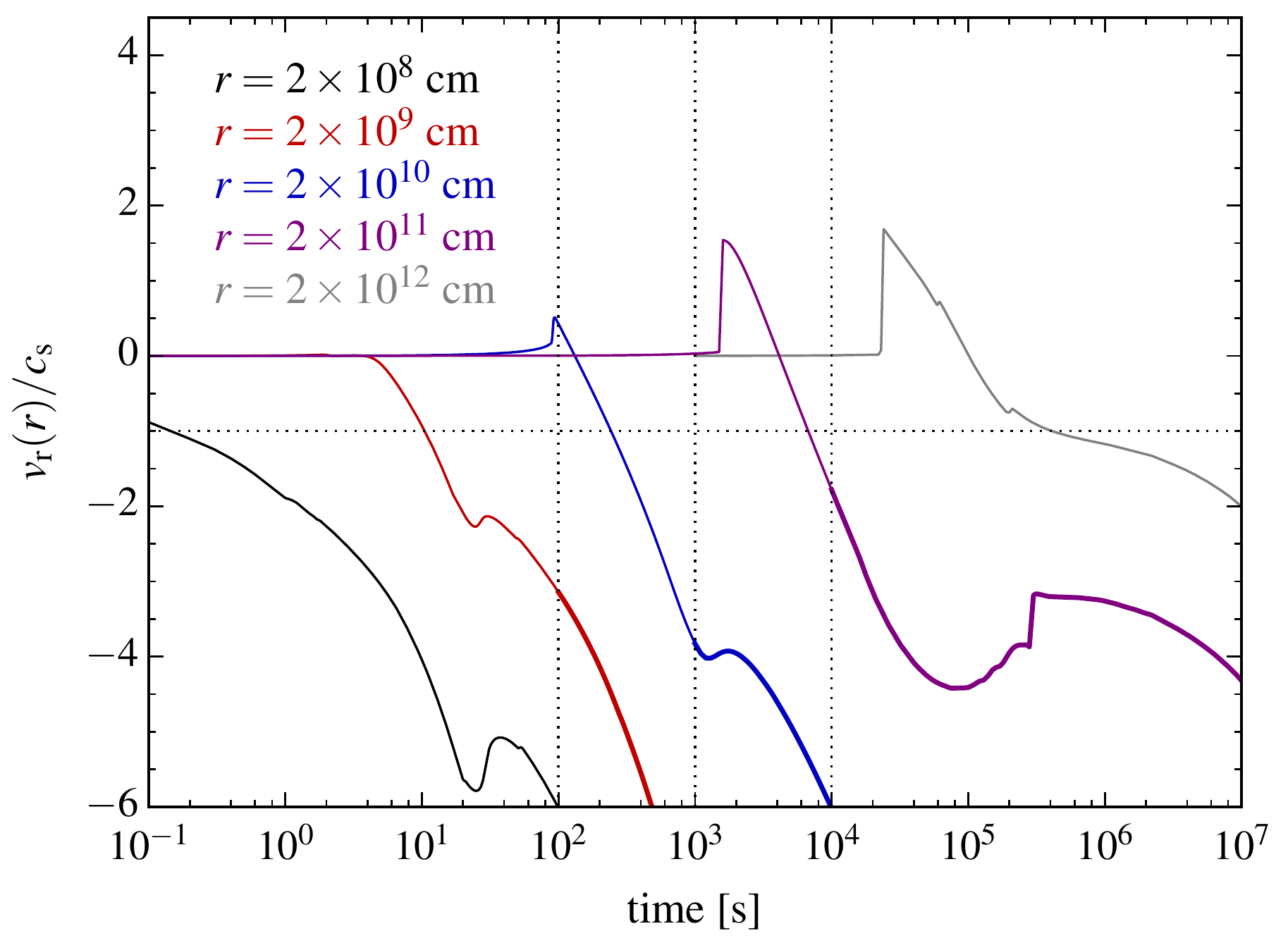}
\caption{Mach number at different radii in the baseline RSG model (M15z00), as labeled. 
The initial inner boundary is located at $r = 2\times 10^8$~cm, and moved out
by a factor of $10$ at the times indicated by the vertical dotted lines, with thick
lines showing the new inner boundary. Curves below the horizontal dotted line are
in supersonic infall.}
\label{f:mach-dn_time_remap}
\end{figure}

Figure~\ref{f:mach-dn_time_remap} shows the Mach number at different radii in the
evolution of the baseline RSG model (M15z00). Within $100$~s of evolution, the
inflow Mach number at the initial inner boundary exceeds $5$, while 
at a radius ten times larger ($2\times 10^9$~cm) the flow is also increasingly 
supersonic. At the times indicated by the vertical dotted lines in Figure~\ref{f:mach-dn_time_remap}, the inner
decade in radius is removed from the computational domain, and the mass contained
in this removed domain is added to both the baryonic and gravitational masses.
We choose the time for the first remapping ($100$~s) so that the TOV mass has
already been reached and no neutrino mass loss is occurring. We have checked
that the subsequent evolution is identical whether this inner decade in radius
is removed or not. The corresponding gains in time step are at least a factor
$10$ in each case, easily allowing evolution of the shock to the surface of the
RSG. No remapping is carried out after $10^4$~s, since strong reverse shocks
cause the infall velocity to decrease in magnitude (while still remaining
supersonic at the inner boundary).

\section{Calculation of the Accretion Rate}
\label{s:accretion_appendix}

Here we provide a derivation of the semi-analytic expression for the accretion
rate in equations~(\ref{eq:tacc_main-text})-(\ref{eq:mdot_main-text}). 
We assume that a given mass shell experiences free-fall
from rest from the time at which the trailing edge of the outgoing shell reaches it,
thus neglecting pressure forces which decrease the infall velocity from the 
free-fall value. Stellar rotation is also neglected.

The time it takes a given  shell at radius $r_0$ to fall to a radius $r$ from rest is \citep{bethe_1990}:
\begin{eqnarray}
t_{\rm fall}(r,r_0) & = & \frac{1}{\sqrt{2GM(r_0)}}\int_r^{r_0} \frac{\totd r^\prime}{\sqrt{1/r^\prime - 1/r_0}}\\
& = & \frac{r_0^{3/2}}{\sqrt{2GM(r_0)}}\int_{r/r_0}^1\,\totd x\, \sqrt{\frac{x}{1-x}}\\
& = & \frac{r_0^{3/2}}{\sqrt{2GM(r_0)}}\left[\frac{\pi}{2}-\arcsin\left(\sqrt{\frac{r}{r_0}}\right)\right.\\
&   & \left.\qquad\qquad + \sqrt{\frac{r}{r_0}\left(1-\frac{r}{r_0}\right)}\right]\\
& \simeq & \frac{r_0^{3/2}}{\sqrt{2GM(r_0)}}\left[\frac{\pi}{2}-\frac{2}{3}\left(\frac{r}{r_0}\right)^{3/2}\right],
\end{eqnarray}
where free-fall motion from rest has been assumed ($\alpha=1$ in \citealt{bethe_1990}), and the latter 
equality is valid for $r \ll r_0$. Note that the mass enclosed by the shell $M(r_0)$ is assumed to remain constant.

Mass conservation in the infalling shell implies
\begin{equation}
\label{eq:infall_mass_conservation}
\rho(r)r^2\totd r = \rho_0(r_0)r_0^2\totd r_0.
\end{equation}
where $\rho_0$ is the density at the time infall begins. For $r \ll r_0$ and
fixed infall time $t_{\rm fall}$, changes in the initial and final radii are related by
\begin{equation}
\label{eq:density_mapping}
\left(\frac{\partial r_0}{\partial r}\right)_t = \frac{4}{3\pi}\left(\frac{r}{r_0} \right)^{1/2},
\end{equation} 
where we have also assumed $\rho_0 r_0^3 \ll M(r_0)$, which is generally true for
shells in the outer envelope of the star ($r_0 \gg r_c$).
Substituting into equation~(\ref{eq:infall_mass_conservation}) yields a relation
between the initial and final densities \citep{bethe_1990}
\begin{equation}
\label{eq:density_infall}
\rho(r) = \frac{4}{3\pi}\,\left(\frac{r_0}{r}\right)^{3/2}\rho_0(r_0).  
\end{equation}

Assuming that the infalling shell has reached the free-fall velocity by the
time it reaches a radius $r$ (an excellent approximation if $r\ll r_0$),
the accretion rate is
\begin{eqnarray}
\label{eq:mdot_appendix}
\dot{M}(r,r_0) & = & f_{\rm fall}4\pi r^2 \rho(r) \sqrt{\frac{2GM(r_0)}{r}}\\
               & = & f_{\rm fall}\frac{16}{3}r_0^2 \rho_0(r_0) \sqrt{\frac{2GM(r_0)}{r_0}}.
\end{eqnarray}
where we have added a fudge factor $f_{\rm fall}$ to correct the density
mapping (equation~\ref{eq:density_mapping}) for the effects of gas pressure
during the early part of the infall\footnote{\citet{bethe_1990} uses a factor $\alpha$ for the
infall velocity to account for the effects of gas pressure.}. 
In practice, we find that setting $f_{\rm fall}=1/2$ 
provides excellent agreement with our simulations (c.f. Figure~\ref{f:mdot_calibration}).

Applying equation~(\ref{eq:mdot_appendix}) to the trailing edge of the
outgoing shock, we have $r_0 = r_{\rm tr}(t)$. The time from the beginning
of the simulation at which the infalling
shell reaches a radius $r$ is
\begin{equation}
\label{eq:tacc_appendix}
t_{\rm acc}(t) = t + t_{\rm fall}(r,r_{\rm tr}[t]),
\end{equation}
which is the time at which the accretion rate in equation~(\ref{eq:mdot_appendix}) is valid.
The very weak dependence of $t_{\rm fall}$ on $r$ for $r\ll r_0$ means that the accretion
rate is primarily a function of time, not of position.


\bibliographystyle{mn2e}
\bibliography{ms}

\label{lastpage}

\end{document}